\pdfoutput=1
\documentclass[12pt,a4paper]{article}

\usepackage{ifthen} 
\newboolean{pdflatex}
\setboolean{pdflatex}{true} 

\newboolean{articletitles}
\setboolean{articletitles}{true} 

\newboolean{uprightparticles}
\setboolean{uprightparticles}{false} 

\usepackage{array} 
\usepackage{booktabs} 
\usepackage{multirow} 

\makeatletter
\g@addto@macro\bfseries{\boldmath}
\makeatother

\def\paperauthors{LHCb collaboration} 
\def\paperasciititle{Study of $B_{c}(1P)^{+}$ states in the $B_{c}^{+} \gamma$ mass spectrum}
\def\paperkeywords{{High Energy Physics}, {LHCb}} 
\def\papercopyright{\the\year\ CERN for the benefit of the LHCb collaboration} 
\def\paperlicence{CC BY 4.0 licence}
\def\paperlicenceurl{https://creativecommons.org/licenses/by/4.0/}

\newif\ifEnableSectionTOCLinks
\EnableSectionTOCLinksfalse 

\usepackage[top=1in, bottom=1.25in, left=1in, right=1in]{geometry}

\usepackage{microtype}
\usepackage{lineno}  
\usepackage{xspace} 
\usepackage{caption} 

\usepackage{graphicx}  
\usepackage{color}
\usepackage{colortbl}
\graphicspath{{./figs/}} 

\usepackage{amsmath} 
\usepackage{amssymb}
\usepackage{amsfonts}
\usepackage{upgreek} 

\newcommand*\patchAmsMathEnvironmentForLineno[1]{%
\expandafter\let\csname old#1\expandafter\endcsname\csname #1\endcsname
\expandafter\let\csname oldend#1\expandafter\endcsname\csname
end#1\endcsname
 \renewenvironment{#1}%
   {\linenomath\csname old#1\endcsname}%
   {\csname oldend#1\endcsname\endlinenomath}%
}
\newcommand*\patchBothAmsMathEnvironmentsForLineno[1]{%
  \patchAmsMathEnvironmentForLineno{#1}%
  \patchAmsMathEnvironmentForLineno{#1*}%
}
\AtBeginDocument{%
\patchBothAmsMathEnvironmentsForLineno{equation}%
\patchBothAmsMathEnvironmentsForLineno{align}%
\patchBothAmsMathEnvironmentsForLineno{flalign}%
\patchBothAmsMathEnvironmentsForLineno{alignat}%
\patchBothAmsMathEnvironmentsForLineno{gather}%
\patchBothAmsMathEnvironmentsForLineno{multline}%
\patchBothAmsMathEnvironmentsForLineno{eqnarray}%
}

\usepackage{hyperxmp}

\usepackage[pdftex,
            pdfauthor={\paperauthors},
            pdftitle={\paperasciititle},
            pdfkeywords={\paperkeywords},
            pdfcopyright={Copyright (C) \papercopyright},
            pdflicenseurl={\paperlicenceurl}]{hyperref}

\usepackage[colorinlistoftodos,textsize=scriptsize]{todonotes}

\usepackage[bottom,flushmargin,hang,multiple]{footmisc}

\usepackage[all]{hypcap} 

\usepackage{xspace} 
\usepackage{upgreek}

\def\lhcb   {\mbox{LHCb}\xspace}

\def\MagUp {\mbox{\em Mag\kern -0.05em Up}\xspace}

\ifthenelse{\boolean{uprightparticles}}%
{

 \def\Ppi         {\ensuremath{\uppi}\xspace}

 \def\Pchi        {\ensuremath{\upchi}\xspace}                 
 \def\Ppsi        {\ensuremath{\uppsi}\xspace}

 \def\PDelta      {\ensuremath{\Delta}\xspace}                 
 \def\PXi         {\ensuremath{\Xi}\xspace}                 
 \def\PLambda     {\ensuremath{\Lambda}\xspace}                 
 \def\PSigma      {\ensuremath{\Sigma}\xspace}                 
 \def\POmega      {\ensuremath{\Omega}\xspace}                 
 \def\PUpsilon    {\ensuremath{\Upsilon}\xspace}
 \let\oldPi\Pi
 \def\PPi         {\ensuremath{\oldPi}\xspace}

 \def\PB      {\ensuremath{\mathrm{B}}\xspace}                 
 \def\PD      {\ensuremath{\mathrm{D}}\xspace}                 
 \def\PJ      {\ensuremath{\mathrm{J}}\xspace}                 
 \def\PK      {\ensuremath{\mathrm{K}}\xspace}                 
 \def\Pb      {\ensuremath{\mathrm{b}}\xspace}                 
 \def\Pc      {\ensuremath{\mathrm{c}}\xspace}

 \def\Pp      {\ensuremath{\mathrm{p}}\xspace}                 

 \def\Ps      {\ensuremath{\mathrm{s}}\xspace}

 \def\thebaroffset{0.0em}
}
{

 \def\Ppi         {\ensuremath{\pi}\xspace}

 \def\Pchi        {\ensuremath{\chi}\xspace}                 
 \def\Ppsi        {\ensuremath{\psi}\xspace}                 
                  
 \mathchardef\PDelta="7101
 \mathchardef\PXi="7104
 \mathchardef\PLambda="7103
 \mathchardef\PSigma="7106
 \mathchardef\POmega="710A
 \mathchardef\PUpsilon="7107
 \mathchardef\PPi="7105
 \def\PB      {\ensuremath{B}\xspace}                 
 \def\PD      {\ensuremath{D}\xspace}                 
 \def\PJ      {\ensuremath{J}\xspace}                 
 \def\PK      {\ensuremath{K}\xspace}                 
 \def\Pb      {\ensuremath{b}\xspace}                 
 \def\Pc      {\ensuremath{c}\xspace}

 \def\Pp      {\ensuremath{p}\xspace}                 

 \def\Ps      {\ensuremath{s}\xspace}

 \def\thebaroffset{0.18em}
}
\newcommand{\offsetoverline}[2][\thebaroffset]{\kern #1\overline{\kern -#1 #2}}%

\makeatletter
\ifcase \@ptsize \relax
  \newcommand{\miniscule}{\@setfontsize\miniscule{4}{5}}
\or
  \newcommand{\miniscule}{\@setfontsize\miniscule{5}{6}}
\or
  \newcommand{\miniscule}{\@setfontsize\miniscule{5}{6}}
\fi
\makeatother

\DeclareRobustCommand{\optbar}[1]{\shortstack{{\miniscule (\rule[.5ex]{1.25em}{.18mm})}
  \\ [-.7ex] $#1$}}

\def\squark    {{\ensuremath{\Ps}}\xspace}

\def\cquark    {{\ensuremath{\Pc}}\xspace}

\def\bquark    {{\ensuremath{\Pb}}\xspace}

\def\pion   {{\ensuremath{\Ppi}}\xspace}
\def\piz    {{\ensuremath{\pion^0}}\xspace}
\def\pip    {{\ensuremath{\pion^+}}\xspace}

\def\KorKbar {\kern \thebaroffset\optbar{\kern -\thebaroffset \PK}{}\xspace}

\def\D       {{\ensuremath{\PD}}\xspace}

\def\DorDbar {\kern \thebaroffset\optbar{\kern -\thebaroffset \PD}\xspace}

\def\Dp      {{\ensuremath{\D^+}}\xspace}
\def\Dm      {{\ensuremath{\D^-}}\xspace}

\def\DpDm    {\ensuremath{\Dp {\kern -0.16em \Dm}}\xspace}

\def\Dsp     {{\ensuremath{\D^+_\squark}}\xspace}

\def\B       {{\ensuremath{\PB}}\xspace}

\def\BorBbar {\kern \thebaroffset\optbar{\kern -\thebaroffset \PB}\xspace}

\def\Bd      {{\ensuremath{\B^0}}\xspace}

\def\BdorBdbar {\kern \thebaroffset\optbar{\kern -\thebaroffset \Bd}\xspace}

\def\Bs      {{\ensuremath{\B^0_\squark}}\xspace}

\def\BsorBsbar {\kern \thebaroffset\optbar{\kern -\thebaroffset \Bs}\xspace}
\def\Bc      {{\ensuremath{\B_\cquark^+}}\xspace}
\def\Bcp     {{\ensuremath{\B_\cquark^+}}\xspace}

\def\jpsi     {{\ensuremath{{\PJ\mskip -3mu/\mskip -2mu\Ppsi}}}\xspace}

\def\chiczero {{\ensuremath{\Pchi_{\cquark 0}}}\xspace}
\def\chicone  {{\ensuremath{\Pchi_{\cquark 1}}}\xspace}
\def\chictwo  {{\ensuremath{\Pchi_{\cquark 2}}}\xspace}

\def\Y#1S{\ensuremath{\PUpsilon{(#1S)}}\xspace}

\def\proton      {{\ensuremath{\Pp}}\xspace}

\def\LorLbar     {\kern \thebaroffset\optbar{\kern -\thebaroffset \PLambda}\xspace}

\def\BF         {{\ensuremath{\mathcal{B}}}\xspace}

\newcommand{\decay}[2]{\mbox{\ensuremath{#1\!\to #2}}\xspace} 

\def\to                 {\ensuremath{\rightarrow}\xspace}

\def\AT#1     {\ensuremath{A_{\mathrm{T}}^{#1}}\xspace}           

\def\C#1      {\ensuremath{\mathcal{C}_{#1}}\xspace}                       
\def\Cp#1     {\ensuremath{\mathcal{C}_{#1}^{'}}\xspace}                    
\def\Ceff#1   {\ensuremath{\mathcal{C}_{#1}^{\mathrm{(eff)}}}\xspace}        
\def\Cpeff#1  {\ensuremath{\mathcal{C}_{#1}^{'\mathrm{(eff)}}}\xspace}       
\def\Ope#1    {\ensuremath{\mathcal{O}_{#1}}\xspace}                       
\def\Opep#1   {\ensuremath{\mathcal{O}_{#1}^{'}}\xspace}                    

\newcommand{\nospaceunit}[1]{\ensuremath{\text{#1}}}       
\newcommand{\aunit}[1]{\ensuremath{\text{\,#1}}}       

\newcommand{\tev}{\aunit{Te\kern -0.1em V}\xspace}
\newcommand{\gev}{\aunit{Ge\kern -0.1em V}\xspace}
\newcommand{\mev}{\aunit{Me\kern -0.1em V}\xspace}
\newcommand{\kev}{\aunit{ke\kern -0.1em V}\xspace}
\newcommand{\ev}{\aunit{e\kern -0.1em V}\xspace}
 
\newcommand{\mevc}{\ensuremath{\aunit{Me\kern -0.1em V\!/}c}\xspace}
\newcommand{\gevc}{\ensuremath{\aunit{Ge\kern -0.1em V\!/}c}\xspace}
\newcommand{\mevcc}{\ensuremath{\aunit{Me\kern -0.1em V\!/}c^2}\xspace}
\newcommand{\gevcc}{\ensuremath{\aunit{Ge\kern -0.1em V\!/}c^2}\xspace}

\def\mum  {\ensuremath{\,\upmu\nospaceunit{m}}\xspace}

\def\fb   {\ensuremath{\aunit{fb}}\xspace}
\def\invfb   {\ensuremath{\fb^{-1}}\xspace}

\def\ps   {\ensuremath{\aunit{ps}}\xspace}

\newcommand{\chisq}{\ensuremath{\chi^2}\xspace}

\newcommand{\chisqip}{\ensuremath{\chi^2_{\text{IP}}}\xspace}

\def\gsim{{~\raise.15em\hbox{$>$}\kern-.85em
          \lower.35em\hbox{$\sim$}~}\xspace}
\def\lsim{{~\raise.15em\hbox{$<$}\kern-.85em
          \lower.35em\hbox{$\sim$}~}\xspace}

\def\sqs   {\ensuremath{\protect\sqrt{s}}\xspace}

\def\pt         {\ensuremath{p_{\mathrm{T}}}\xspace}

\def\ptot       {\ensuremath{p}\xspace}

\newcommand{\lum} {\ensuremath{\mathcal{L}}\xspace}

\def\evtgen     {\mbox{\textsc{EvtGen}}\xspace}

\def\geant      {\mbox{\textsc{Geant4}}\xspace}

\def\photos     {\mbox{\textsc{Photos}}\xspace}

\def\pythia     {\mbox{\textsc{Pythia}}\xspace}

\def\tell1  {TELL1\xspace}
\def\ukl1   {UKL1\xspace}

\newcommand{\lhcborcid}[1]{\href{https://orcid.org/#1}{\hspace*{0.1em}\raisebox{-0.45ex}{\includegraphics[width=1em]{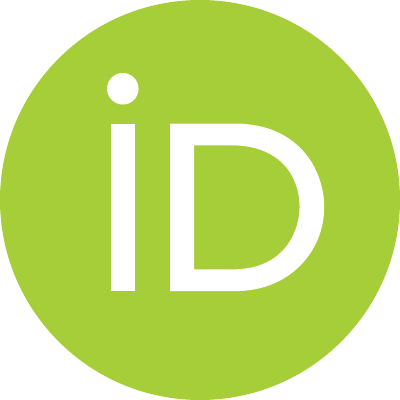}}}}

\usepackage{cite} 
\usepackage{mciteplus}

\usepackage{longtable} 

\newcommand{\xx}{\ensuremath{\kern 0.5em }}

\def\Bcstar    {{\ensuremath{\B_\cquark^{*+}}}\xspace}

\usepackage{multirow} 
\usepackage{booktabs} 
\usepackage{rotating}

\def\BcOneP   {B_{c}(1P)^{+}}
\def\BcGamma   {B_{c}^{+}\gamma}
\def\BcToJpsipi  {\decay{\Bcp}{\jpsi\pi^{+}}}
\def\BcOnePToBc  {\decay{\BcOneP}{\Bcp\gamma}}

\def\chicone {\chi_{c1}}

\def\chictwo {\chi_{c2}}

\def\ntracks {n_{\text{tracks}}}

\def\Dsst {D_{s1}(2460)^{+}}
\def\Dsp {D_{s}^{+}}

\def\degree{\ensuremath{^{\circ}}\xspace}

\usepackage{placeins}

\begin{document}

\renewcommand{\thefootnote}{\fnsymbol{footnote}}
\setcounter{footnote}{1}


\begin{titlepage}
\pagenumbering{roman}

\vspace*{-1.5cm}
\centerline{\large EUROPEAN ORGANIZATION FOR NUCLEAR RESEARCH (CERN)}
\vspace*{1.5cm}
\noindent
\begin{tabular*}{\linewidth}{lc@{\extracolsep{\fill}}r@{\extracolsep{0pt}}}
\ifthenelse{\boolean{pdflatex}}
{\vspace*{-1.5cm}\mbox{\!\!\!\includegraphics[width=.14\textwidth]{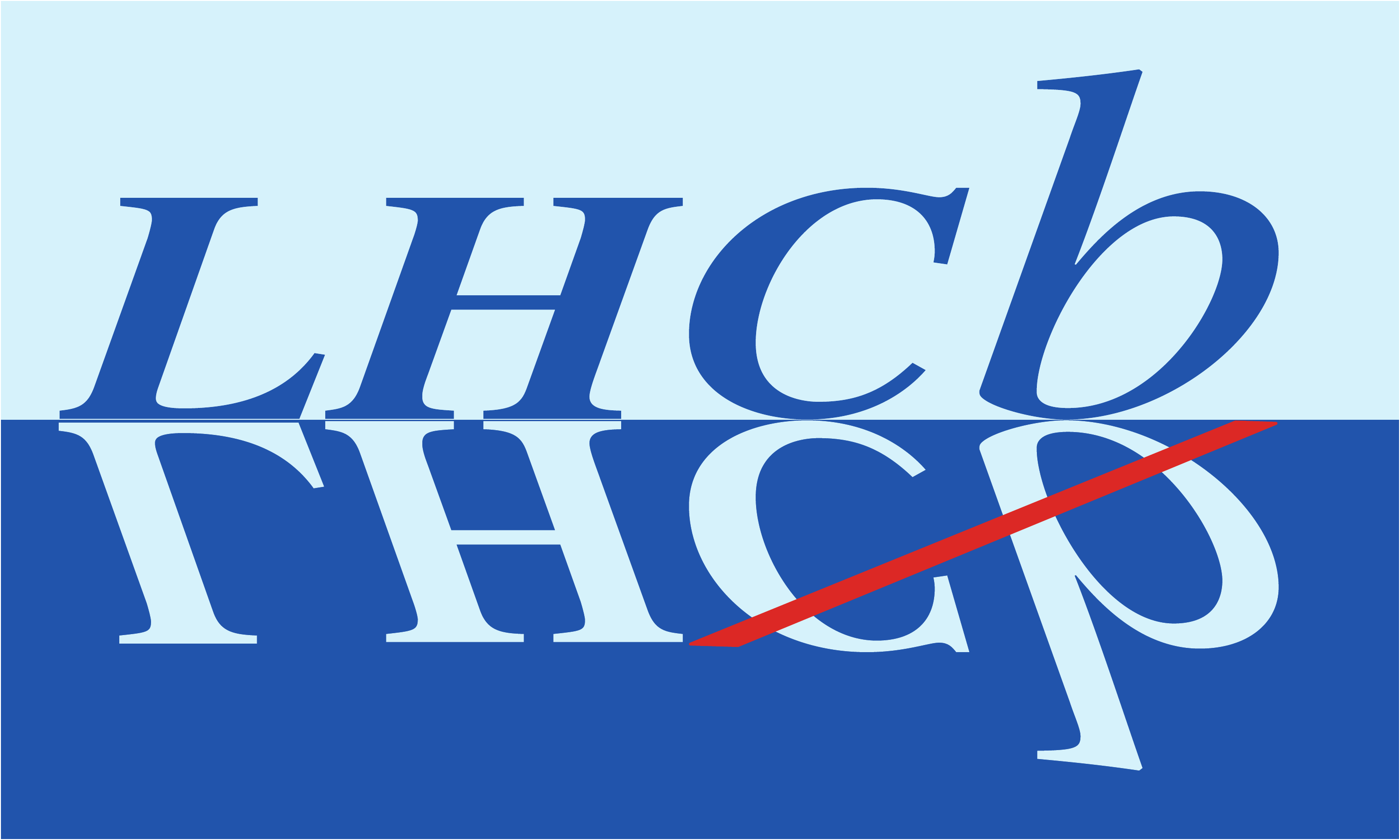}} & &}%
{\vspace*{-1.2cm}\mbox{\!\!\!\includegraphics[width=.12\textwidth]{figs/lhcb-logo.eps}} & &}%
\\
 & & CERN-EP-2025-122 \\  
 & & LHCb-PAPER-2025-015 \\  
 & & 3 December 2025 \\ 
 & & \\
\end{tabular*}

\vspace*{4.0cm}

{\normalfont\bfseries\boldmath\huge
\begin{center}
Study of $B_c(1P)^+$ states in the $B_c^+ \gamma$ mass spectrum
\end{center}
}

\vspace*{1.0cm}

\begin{center}
\paperauthors\footnote{Authors are listed at the end of this paper.}
\end{center}

\vspace{\fill}

\begin{abstract}
  \noindent
    The study of a wide peaking structure in the $B_{c}^{+} \gamma$ mass spectrum is reported using a data sample of proton-proton collisions collected by the LHCb detector at center-of-mass energies of $7$, $8$, and $13~\text{TeV}$, corresponding to an integrated luminosity of $9~\text{fb}^{-1}$.
    The observed structure is consistent with the lowest excited $P$-wave $B_{c}^{+}$ states and exhibits a statistical significance exceeding 7 standard deviations relative to the background-only hypothesis.
    A two-peak model serves as an effective description of the data, with various theory-constrained models further explored to provide physical interpretation.
    Based on the predictions for the $B_{c}(1P)^{+}$ spectrum, 
    the relative production cross section of the overall $B_{c}(1P)^{+}$ states with respect to the $B_{c}^{+}$ ground state with the transverse momentum $p_{\text{T}}$ and rapidity $y$ of $B_{c}^{+}$ mesons in the regions $p_{\text{T}}<20~\mathrm{Ge\kern -0.1em V\!/}c$ and $2.0<y<4.5$ at $\sqrt{s}=13~\text{TeV}$ is measured to be $0.20 \pm 0.03 \pm 0.02 \pm 0.03$, where the uncertainty terms represent statistical, systematic, and uncertainties related to the choice of theoretical models, respectively.
    The results provide a test of theoretical models and deepen our understanding of quantum chromodynamics.
\end{abstract}

\vspace*{1.0cm}

\begin{center}
  Published in Phys. Rev. D 112 (2025) 112003
\end{center}

\vspace{\fill}

{\footnotesize 
\centerline{\copyright~\papercopyright. \href{\paperlicenceurl}{\paperlicence}.}}
\vspace*{2mm}

\end{titlepage}


\newpage
\setcounter{page}{2}
\mbox{~}
%
%
%
%

\onecolumn

\renewcommand{\thefootnote}{\arabic{footnote}}
\setcounter{footnote}{0}

\cleardoublepage

\pagestyle{plain} 
\setcounter{page}{1}
\pagenumbering{arabic}



\section{Introduction}
\label{sec:introduction}

The $\Bcp$ meson family, composed uniquely of two different heavy quarks, serves as a distinctive probe of the nonperturbative regime of quantum chromodynamics (QCD).
It is predicted to have a rich spectrum, which serves as a crucial test bed for theoretical calculations based on lattice QCD~\cite{Davies:1996gi,Shanahan:1999mv,Allison:2004be,Mathur:2018epb} and various other QCD-based calculations~\cite{
Gershtein:1994dxw,Gupta:1995ps,Fulcher:1998ka,Ebert:2002pp,Godfrey:2004ya,Eichten:2019gig,Li:2019tbn,Wang:2022cxy,Li:2022bre,Li:2023wgq,Hao:2024nqb,
Chen:1992fq, Zeng:1994vj, El:2005, Wei:2010zza, Dowdall:2012ab, Wang:2012kw,
Karliner:2018bms,
Chen:2020ecu, Martin-Gonzalez:2022qwd,Nussinov:1983hb,Bertlmann:1979zs,Soni:2017wvy},
thereby advancing our understanding of the dynamical mechanisms governing the hadronic structure.

Since the observation of the $\Bcp$ ground state~\footnote{Charge conjugation is implied throughout the paper.} by the CDF experiment at the Tevatron collider~\cite{Abe:1998wi,Abe:1998fb},
the investigation of the $\Bcp$ spectrum has long been an experimental pursuit.
The excited $\Bcp$ states with masses below the $BD$ threshold, which corresponds to the sum of the masses of a beauty and a charmed meson, can only decay to the ground state through radiative transitions or pion emission~\cite{Gouz:2002kk, Godfrey:2004ya,Eichten:2019gig}.
Therefore, the $\Bcp\gamma$ and $B_c^+\pi^+\pi^-$ systems represent the only channels to identify these low-lying excited $\Bcp$ states.
However,
these searches are inherently challenging 
since the $\Bcp$ mesons are beyond the reach of currently operating $e^+e^-$ collider facilities,
and their production in hadron collisions requires 
simultaneous $c\bar{c}$ and $b\bar{b}$ pair production and is therefore relatively small~\cite{Eichten:2019gig,Chang:2005bf}.
The operation of the LHC at unprecedented center-of-mass energies and high luminosity has revolutionized studies of the $\Bcp$ mesons.
The second $S$-wave excited state, $B_c(2S)^+$, was first reported by the ATLAS experiment as a wide structure in the $B_c^+\pi^+\pi^-$ mass spectrum~\cite{Aad:2014laa}, 
which was later resolved into two narrow peaks corresponding to the $B_c(2^1S_0)^+$ and $B_c(2^3S_1)^+$ states~\footnote{
Here applies the spectroscopic notation, $n^{2S+1}L_{J}$, where $n$ gives the radial quantum number, $S$ the sum of the quark spins, $L$ the orbital angular momentum of the quarks, and $J$ the total angular momentum.}
by the CMS and LHCb experiments~\cite{CMS:2019uhm,LHCb-PAPER-2019-007}.
To probe the $B_c(1P)^+$ states, a study of the $\Bcp\gamma$ mass spectrum is necessary. 
The LHCb experiment is uniquely positioned for that purpose,
as evidenced by several measurements of the mass, lifetime, production and decay modes of the $B_c^+$ states, summarized in Ref.~\cite{LHCb-PAPER-2023-046}, 
and successful experience with the radiative decays of quarkonia~\cite{
LHCb-PAPER-2011-019,LHCb-PAPER-2011-030,LHCb-PAPER-2013-028,
LHCb-PAPER-2014-040,LHCb-PAPER-2014-031}.

The combination between the total spin ($S=0$ or $1$) of the $\bar{b}c$ system and the orbital angular momentum ($L=1$) between the two quarks
gives rise to four $B_c(1P)^+$ states:
the spin singlet $1^1P_1$ and the spin triplet $1^3P_0$, $1^3P_1$ and $1^3P_2$.
The $1^1P_1$ state decays to the $\Bcp\gamma$ final state,
while the spin triplet states decay to $B_{c}(1^3S_1)^+$ (denoted as $\Bcstar$ hereafter), which subsequently decays to the $\Bcp\gamma$ final state~\cite{Gouz:2002kk, Godfrey:2004ya,Eichten:2019gig}.
The photon originating from the $\Bcstar$ decay is expected to carry such a small energy
that it escapes detection~\cite{Berezhnoy:2013sla,Berezhnoy:2019yei} and
the reconstructed $\Bcp\gamma$ mass distribution of the $B_c(1P)^+$ states shows a peaking structure shifted to lower values, where the shift is determined by the mass difference $\delta M$ between the $\Bcstar$ and $\Bcp$ states.
Furthermore, the $1^1P_1$ and $1^3P_1$ states mix into the physical states $1P'_1$ and $1P_1$ according to
\begin{equation}
\label{eq:mixing}
\begin{pmatrix}
1P'_1 \\
1P_1
\end{pmatrix}
=
\begin{pmatrix}
\phantom{-}\cos\theta & \sin\theta \\
-\sin\theta & \cos\theta
\end{pmatrix}
\begin{pmatrix}
1^1P_1 \\
1^3P_1
\end{pmatrix},
\end{equation}
with $\theta$ the mixing angle.
Due to mixing, both physical states $1P'_1$ and $1P_1$ decay to the $\Bcp\gamma$ and $\Bcstar\gamma$ final states.
Their masses, as well as those of the $1^3P_0$ and $1^3P_2$ states, are calculated in Refs.~\cite{
Davies:1996gi, 
Gershtein:1994dxw,Gupta:1995ps,Fulcher:1998ka,Ebert:2002pp,Godfrey:2004ya,Eichten:2019gig,Li:2019tbn,Wang:2022cxy,Li:2022bre,Li:2023wgq,Hao:2024nqb} using lattice QCD and 
numerous theory approaches, with these selected predictions summarized in Table~\ref{tab:mass_predictions}.
The most precise $\Bcp$ mass predictions within the framework of lattice QCD have been provided by a recent study\mbox{~\cite{Mathur:2018epb}}\hskip0pt, albeit for a subset of the $\BcOneP$ states.
In total, six peaks from $B_c(1P)^+$ states are expected in the $\Bcp\gamma$ mass spectrum: Two correspond to the masses of the $1P_1$ and $1P_1'$ states, while the remaining four are displaced by $\delta M$ from the masses of the $1^3P_0$, $1P_1$, $1P_1'$, and $1^3P_2$ states.
The differences between the reconstructed $\BcOneP$ masses in the $\Bcp\gamma$ mass spectrum and the ground-state $\Bcp$ mass,
denoted as $M(\Bcp \gamma)-M(\Bcp)$,
are expected to be in the range $[340,520] \mevcc$~\cite{
Davies:1996gi, 
Gershtein:1994dxw,Gupta:1995ps,Fulcher:1998ka,Ebert:2002pp,Godfrey:2004ya,Eichten:2019gig,Li:2019tbn,Wang:2022cxy,Li:2022bre,Li:2023wgq,Hao:2024nqb}.

\begin{table}[t]
\begin{center}
\caption{Predicted masses for the $1P$ $\Bcp$ states and the mass difference \mbox{$\delta M = M(\Bcstar) - M(\Bcp)$} between the two first $S$-wave states (in units of \mevcc) , along with the mixing angle $\theta$.}
\label{tab:mass_predictions}
\begin{tabular}{l|c|cccc|c}
\toprule
         & $\delta M$ & $M(1^3P_0)$ & $M(1P_1)$ & $M(1P_1')$ & $M(1^3P_2)$ & $\quad\theta$ [$\degree$] \\
\midrule
Lattice QCD~\cite{Davies:1996gi}      & 41     & 6727   & 6743   & 6765   & 6783   & \phantom{$-$}$33.4$ \\
GKLT~\cite{Gershtein:1994dxw}  & 64     & 6683   & 6717   & 6729   & 6743   & \phantom{$-$}$17.1$ \\
GJ~\cite{Gupta:1995ps}           & 61     & 6689   & 6738   & 6757   & 6773   & \phantom{$-$}$25.6$ \\
FUII~\cite{Fulcher:1998ka}       & 55     & 6701   & 6737   & 6760   & 6772   & \phantom{$-$}$28.5$ \\
EFG~\cite{Ebert:2002pp}              & 62     & 6699   & 6734   & 6749   & 6762   & \phantom{$-$}$20.4$ \\
GI~\cite{Godfrey:2004ya}          & 67     & 6706   & 6741   & 6750   & 6768   & \phantom{$-$}$22.4$ \\
EQ~\cite{Eichten:2019gig}        & 54     & 6693 & 6731 & 6739 & 6759 & \phantom{$-$}$18.7$ \\
LLLLGZ~\cite{Li:2019tbn}                    & 55     & 6714   & 6757   & 6776   & 6787   & \phantom{$-$}$35.5$ \\
WWLC~\cite{Wang:2022cxy}           & 55     & 6705   & 6739   & 6748   & 6762   & \phantom{$-$}$32.2$ \\
LTFWP~\cite{Li:2022bre}                 & 53     & 6712   & 6770   & 6761   & 6783   & $-24.3$ \\
LLWL~\cite{Li:2023wgq}               & 67     & 6701   & 6745   & 6754   & 6773   & \phantom{$-$}$35.2$ \\
HZ~\cite{Hao:2024nqb}                 & 63     & 6707   & 6751   & 6786   & 6802   & \phantom{$-$}$55.0$   \\
\bottomrule
\end{tabular}
\end{center}
\end{table}

Using proton-proton ($pp$) collision data recorded by the LHCb experiment at center-of-mass energies of $\sqs = 7$, $8$, and $13\tev$,
corresponding to integrated luminosities of $1$, $2$, and $6\invfb$, respectively,
a wide peaking structure consistent with the $B_c(1P)^+$ states is observed in the $\Bcp\gamma$ system reconstructed through the $\Bcp \to \jpsi(\to \mu^+ \mu^-) \pip$ decay chain, as reported in Ref.~\cite{LHCb-PAPER-2025-014}.
This paper focuses on the interpretation of the $B_c(1P)^+$ structure and its production measurement and provides details of the selection criteria and the photon energy correction method.
A brief introduction to the \lhcb detector along with the reconstruction and simulation procedures is provided in Sec.~\ref{sec:detector}. 
The event selection criteria are detailed in Sec.~\ref{sec:selection}. 
The correction of the photon energy is discussed in Sec.~\ref{sec:Ecorrection}. 
The $\BcOneP$ signal yield and corresponding peak locations are determined in Sec.~\ref{sec:massfit} using a theory-independent approach. 
Tests of theoretical models and measurement of $\BcOneP$ production are reported in Sec.~\ref{sec:production}. 
Systematic uncertainties on the production rates are evaluated in Sec.~\ref{sec:systematics}.
All results are summarized in Sec.~\ref{sec:summary}.

\section{Detector and simulation}
\label{sec:detector}

The \lhcb detector \cite{LHCb-DP-2008-001,LHCb-DP-2014-002} is a single-arm forward
spectrometer covering the pseudorapidity range $2<\eta <5$,
designed for the study of particles containing \bquark or \cquark
quarks. 
The detector includes a high-precision tracking system
consisting of a silicon\nobreakdash-strip vertex detector surrounding 
the $\proton\proton$ interaction region, 
a large-area silicon-strip detector located
upstream of a dipole magnet with a bending power of about
$4{\mathrm{\,T\, m}}$, and three stations of silicon\nobreakdash-strip detectors and straw
drift tubes placed downstream of the magnet.
The tracking system provides a measurement of the momentum, \ptot, 
of charged particles with
a relative uncertainty that varies from 0.5\% at low momentum to 1.0\% at 200\gevc.
The minimum distance of a track to 
a primary $\proton\proton$ collision vertex\,(PV), 
the impact parameter\,(IP), 
is measured with a resolution of $(15+29/\pt)\mum$,
where \pt is the component of the momentum transverse to the beam, in\,\gevc. 
Particle identification (PID) for charged hadrons uses information from two ring-imaging
Cherenkov detectors. 
Muons are identified by a system composed of 
alternating layers of iron and multiwire
proportional chambers.

A calorimeter system consisting of
scintillating-pad~(SPD) and preshower detectors, an electromagnetic
calorimeter~(ECAL) and a hadronic calorimeter is responsible for the identification of photons, electrons and hadrons.
The ECAL employs the Shashlik technology, i.e., a sampling scintillator/lead structure readout by plastic wavelength shifting fibers, with a design energy resolution of $\sigma_{E}/E=10\%/\sqrt{E}\oplus1\%$ , with \mbox{$E$ in {Ge\kern-0.1em V}}~\cite{LHCb-TDR-002}.
The ECAL energy scale is calibrated using photons from $\pi^0$ decays, as detailed in Refs.~\cite{Belyaev:2014zga,LHCb-DP-2020-001}.
Photons are reconstructed from the energy deposits in the ECAL that are not associated with any reconstructed tracks~\cite{LHCb-2003-091}. 
For offline calibration purposes, they are separated into two categories based on their response in the SPD detector: those that have converted in the material following the dipole magnet (converted photons) thus leaving hits in the SPD, and those that have not undergone conversion (nonconverted photons) and so do not interact with the SPD.
Photons that convert before the magnet can be reconstructed as a $e^+e^-$ pair.
They have excellent energy resolution  given the accurate tracking information available~\cite{LHCb-PAPER-2014-040,LHCb-PAPER-2013-028}. 
However, the reconstruction efficiency is very low, and therefore, they are not used in this analysis.

The online event selection is performed by a trigger, 
which consists of a hardware stage based on information from the calorimeter and muon
systems, followed by a software stage which performs a full event
reconstruction.
The trigger selection algorithms are primarily based on identifying the key characteristics of beauty and charm hadrons to ensure maximum efficiency across the full spectrum of their possible decays.
This includes considering the topological signature of their displaced vertices and imposing kinematic requirements such as on the transverse momentum of final-state particles. 
At the hardware stage, events are required to have at least one muon with high transverse
momentum or a hadron with high transverse energy. At the software stage, two muon tracks or three charged tracks are required to have high $\pt$ and to form a secondary vertex with a significant displacement from the interaction point.

Simulation is required to model the signal, as well as to study the effects of detector acceptance, resolution, and the selection requirements. 
The {\textsc{BcVegPy}} generator~\cite{Chang:2015qea} is employed to simulate the production of $\BcOneP$ and $\Bcp$ mesons. 
It is based on full perturbative QCD calculations at the lowest order ($\upalpha_{\mathrm{s}}^4$) via the dominant gluon-gluon fusion process, 
while neglecting contributions from the quark-pair annihilation 
~\cite{Chang:1992jb, Chang:1994aw, Chang:1996jt, Chang:2004bh, Berezhnoy:1996ks}. 
The generator is interfaced with the \pythia parton shower and hadronization model~\cite{Sjostrand:2007gs} with a specific LHCb configuration~\cite{LHCb-PROC-2010-056}.
The decays of unstable particles are modeled using \evtgen~\cite{Lange:2001uf}, and final-state radiation is generated using \photos~\cite{davidson2015photos}. 
The interaction of the generated particles with the detector, and its response,
are implemented using the \geant
toolkit~\cite{Allison:2006ve, *Agostinelli:2002hh} as described in
Ref.~\cite{LHCb-PROC-2011-006}. 
The simulated sample is also corrected to match 
the track reconstruction performance~\cite{LHCb-DP-2013-002} and PID response~\cite{LHCb-DP-2013-001} in data.
The detector response used for particle identification is sampled from the $D^{*+} \to D^0 (\to K^- \pi^+) \pi^+$ and $K_S^0 \to \pi^+ \pi^-$ control channels for pions, and the $J/\psi \to \mu^+ \mu^-$ control channel for muons, respectively~\cite{LHCb-DP-2018-001}.
The difference of the photon reconstruction efficiency between data and simulation
is corrected using the relative yields of reconstructed $B^+ \to \jpsi K^{*+} (\to K^+ \piz (\to \gamma \gamma))$ and $B^+ \to \jpsi K^+$ decays~\cite{Govorkova:2015vqa}.
The energy response to photons in simulation is corrected using large calibration samples of $\chicone \to J/\psi \gamma$ decays, with converted and nonconverted photons treated separately, as described in Sec~\ref{sec:Ecorrection}.

\section{Candidate reconstruction and selection}
\label{sec:selection}

The search for $\BcOneP$ states is performed in the decay chain $\BcOneP \to \Bcp\gamma$, with $\Bcp \to \jpsi \pip$ and $\jpsi \to \mu^+ \mu^-$.
The selection procedure of $\BcToJpsipi$ candidates follows that used in Refs.~\cite{LHCb:2020ayi,LHCb:2019bem} as described below. 
Candidate $\jpsi$ mesons are formed by combining pairs of oppositely charged muons,
each required to have $\pt > 550\mevc$, that originate from a common vertex. 
The $\jpsi$ candidates must have a mass within the range $[3040,3140]\mevcc$, which corresponds to approximately six times the $\jpsi$ mass resolution,
and are then combined with a charged pion to form the $\Bcp$ candidates. 
The pion must satisfy $\pt > 1000\mevc$ and be inconsistent with originating from any PV. 
The $\Bcp$ candidates are required to have a good-quality vertex fit, a decay time exceeding $0.2\ps$, and a momentum vector aligned with the displacement vector identified by the decay vertex and the associated PV.
The latter represents the vertex with the smallest impact parameter $\chisqip$, defined as the difference in the vertex-fit $\chi^2$ of a given PV
with and without the $\Bcp$ candidate. 

To further suppress combinatorial background, a boosted decision tree (BDT)~\cite{Breiman,AdaBoost} classifier is applied. 
The input variables comprise the transverse momentum $\pt$ of the muons, the pion, and the $J/\psi$ meson; the decay length, decay time, and vertex-fit $\chisq$ of the $\Bcp$ meson, as well as the $\chisqip$ values of the muons, the pion, the $\jpsi$ meson, and the $\Bcp$ meson relative to the associated PV. 
The classifier is trained on simulated signal candidates and background candidates from the upper sideband of the $\jpsi\pip$ mass spectrum in data, within the mass range $[6370,6600]\mevcc$.  
The BDT selection working point is optimized to maximize the figure of merit $S/\sqrt{S + B}$, 
where $S$ and $B$ denote the expected signal and background yields, respectively, in the mass window $M(\jpsi\pip)\in[6251,6301]\mevcc$, which corresponds to approximately four times the $\Bcp$ mass resolution.

To form the $\BcOneP$ candidates, the selected $\Bcp$ candidates within the mass range $[6240, 6300] \mevcc$ are combined with photons detected by the calorimeter system.
The majority of photons originate from the decay of neutral pions, i.e., $\pi^0 \to \gamma\gamma$. 
This background contribution is reduced by discarding photons with a companion photon such that the diphoton mass is within a $\pm 30 \mevcc$ region from the known $\pi^0$ mass~\cite{PDG2024}.
Further photon selections are implemented to maximize the Punzi figure of merit~\cite{Punzi:2003bu}, $S/(5/2 + \sqrt{B})$, for $\BcOneP$ signals in the expected mass region. These enforce that photons  have transverse energy $E_{\rm T} > 1000 \mev$, $E > 4000 \mev$, 
and a high probability of being correctly identified, based on the value of a neural-net estimator that combines information from the calorimeter and tracking systems~\cite{Hoballah:2015sia}. 
Finally, to reduce background contributions from multiple photon combinations with the same $\Bcp$ meson, only the candidate with the highest $E_{\rm T}$ photon is kept.

\begin{figure}[t]
 \begin{center}
    \includegraphics[width=0.5\textwidth]{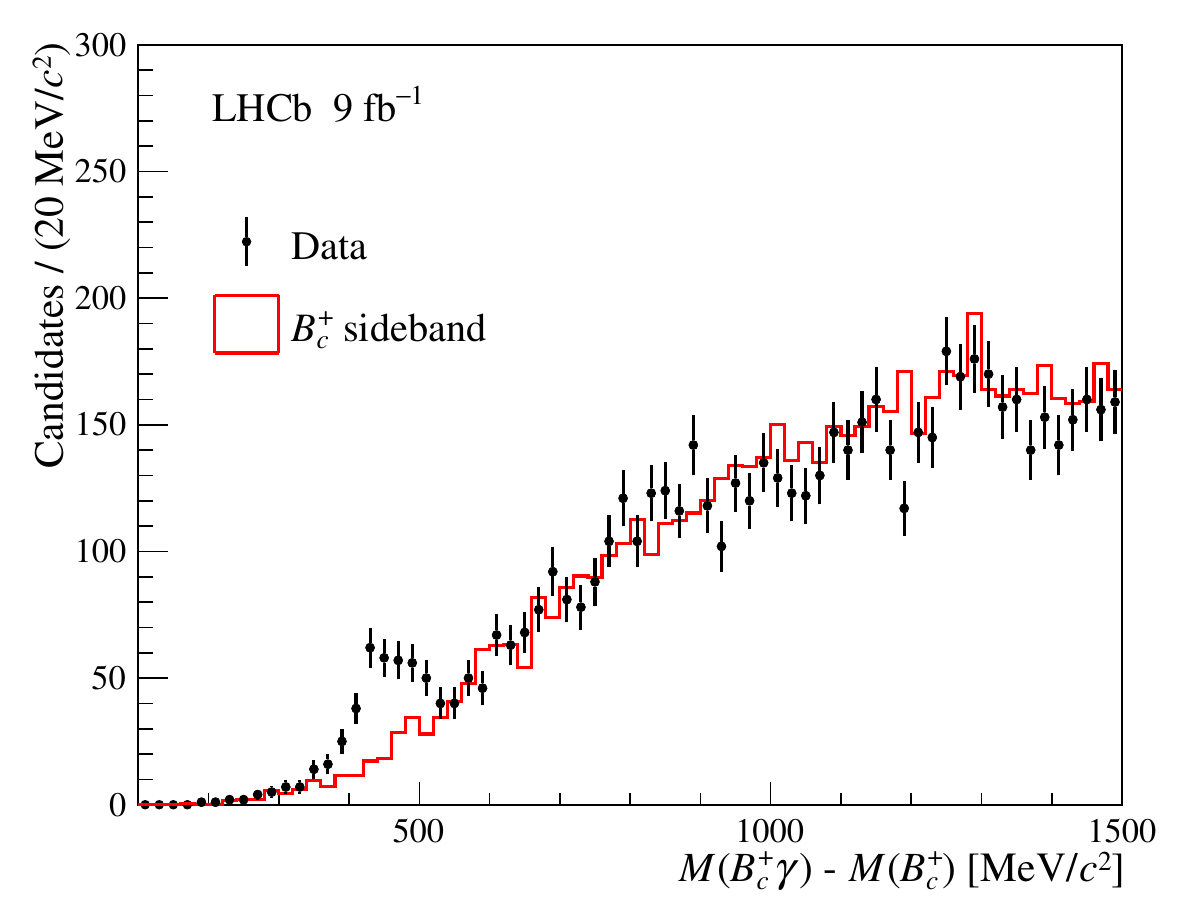}
    \caption{Distribution of $M(B_{c}^{+}\gamma)-M(B_{c}^{+})$ for the selected $B_{c}^{+}$ candidates. The distribution of $B_{c}^{+}$ candidates from the high-mass sideband region is also shown for comparison.}
   \label{fig:mass_comp}
 \end{center}
\end{figure}

The distribution of $M(\Bcp\gamma) - M(\Bcp)$ for selected $\Bcp$ candidates is shown in Fig.~\ref{fig:mass_comp},
This variable is chosen to minimize effects from the $\Bcp$ mass resolution. 
The corresponding  distribution, obtained by considering $\Bcp$ candidates from the high-mass sideband region ($M(\Bcp) \in [6340, 6550] \mevcc$), is also shown for comparison. 
It demonstrates that the chosen selection criteria do not introduce artificial peaks in the pure background sample.
A pronounced peaking structure is evident within the predicted mass range of the $\BcOneP$ states.

\section{Photon energy correction in simulation}
\label{sec:Ecorrection}

To determine the signal composition of the peaking structure, 
a good understanding of the photon energy scale and resolution is required.
Dedicated corrections are applied to the simulation to accurately model the location and width of the signal peaks.
These corrections are obtained using large data and simulated samples of $\chi_{c} \rightarrow J/\psi \gamma$ decays, where the photon energy spectrum is similar to that in the signal decays.
The selection criteria applied to photons are identical to those used for the $\BcOneP$ candidates, which ensures portability of the corrections.
  
\begin{figure}[t]
 \begin{center}
    \includegraphics[width=0.49\textwidth]{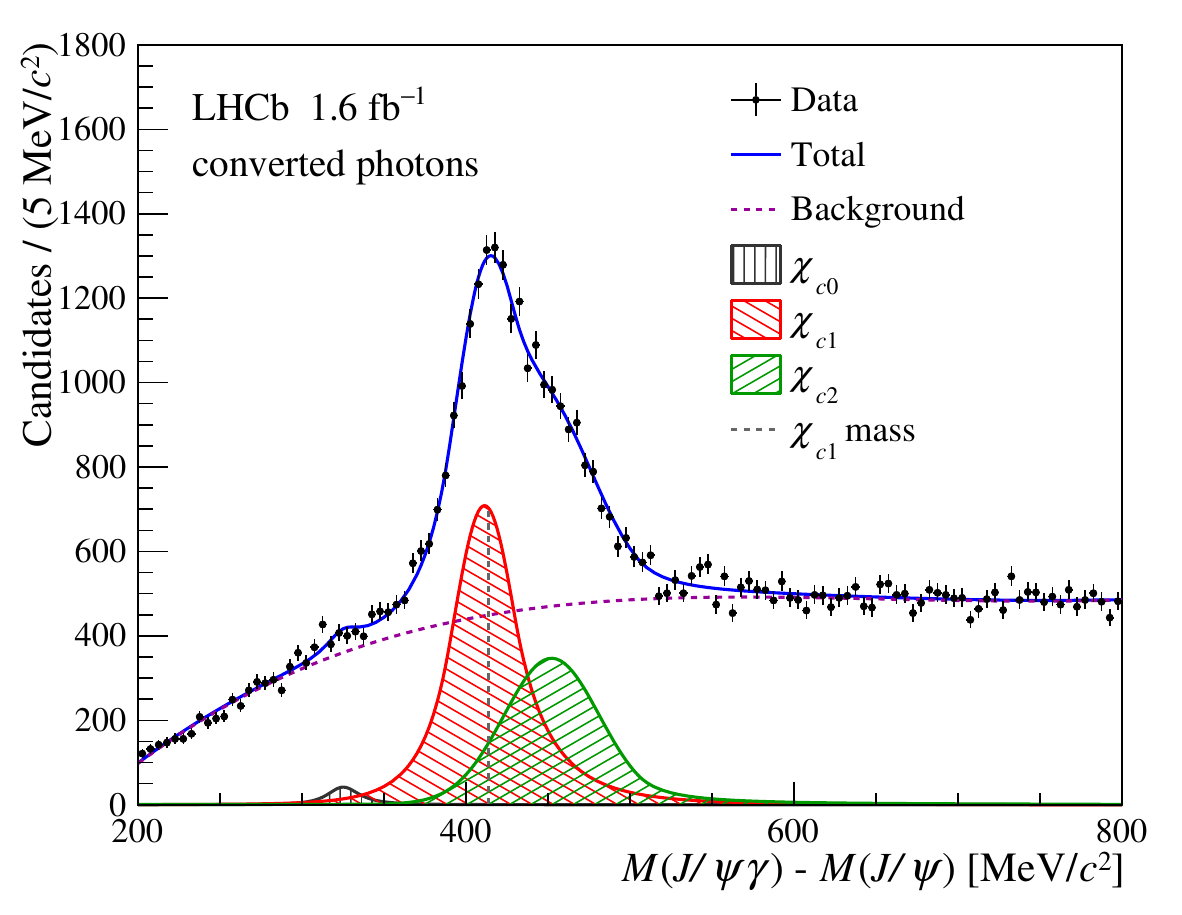}
    \caption{Distribution of $M(J/\psi\gamma) - M(J/\psi)$ for selected $\chi_c$ candidates with converted photons, in the region $E_\gamma \in [5.9, 7.2]~\mathrm{Ge\kern -0.1em V}$ and $n_{\text{tracks}} \in [125, 181]$. Fit results are also shown. Contributions from $\chi_{c0}, \chi_{c1}$, and $\chi_{c2}$ decays are shown with different line colors, while the gray vertical line indicates the known $\chi_{c1}$ mass~\cite{PDG2024}.}
   \label{fig:calib_chicM}
 \end{center}
\end{figure}

Photon energy corrections in simulation are derived
from fits to the $J/\psi\gamma$ mass distributions in data.
They are obtained separately for converted and nonconverted photons in each data taking period,
and in two-dimensional regions of the photon energy, $E_\gamma$, and the number of reconstructed tracks, $\ntracks$.
An example of the $M(\jpsi\gamma) - M(\jpsi)$ distribution of converted photons in one typical $E_\gamma$ and $\ntracks$ region using 2016 data is shown in Fig.~\ref{fig:calib_chicM}.
The distribution shows a wide peaking structure due to the contributions from the $\chiczero$, $\chicone$, and $\chictwo$ states. 
These resonant contributions are each modeled as a nonrelativistic Breit--Wigner function convolved with a modified Gaussian function with asymmetric power-law tails to account for the mass resolution.
The natural widths of the $\chi_c$ states are fixed to their known values~\cite{PDG2024}. 
The background component is modeled by a third-order polynomial function.
In each $E_\gamma$ and $\ntracks$ region, the photon energy bias and resolution are determined based on their linear relationship with two observables: 
the deviation of the $\chi_{c1}$ peak location from its known value~\cite{PDG2024}, and the measured width of the $\chi_{c1}$ mass peak.
The linear dependencies are well established by simulation studies.

The simulated samples of $\BcOneP \to B_c^+\gamma$ are generated with different $\BcOneP$ masses within the theoretical prediction range. The signal shape in $M(B_c^+\gamma)-M(B_c^+)$ distribution is corrected on a per-candidate basis,
using the photon energy bias and resolution of the corresponding $E_\gamma$ and $\ntracks$ region.
With the corrections applied, 
the $B_c^+\gamma$ signal resolution is parametrized as a modified Gaussian function with asymmetric power-law tails.
The resolution, $\sigma$, and peak location, $\mu$, are found to have a small linear dependency on the true mass difference of the $\BcOneP$ signal peak in simulation, $\Delta M_{\rm true}$,
given by $\mu = -0.006 \times \Delta M_{\rm true} -0.340 \mevcc$
and $\sigma = 0.041 \times \Delta M_{\rm true} +1.640 \mevcc$.
These dependencies are used to fully constrain the signal shape for a given mass hypothesis.

A self-consistency check of the photon energy correction is performed by applying the same method to the $\chi_{c1} \rightarrow J/\psi \gamma$ and $\chi_{c2} \rightarrow J/\psi \gamma$ decays.
A good agreement between the corrected mass and the value in data is obtained and the measured mass resolution is well reproduced.
As an example, using samples with converted photons from the 2016 data-taking period, 
the difference between the $\chi_{c1}$ mass determined from the simulation and the data is reduced from $2.0\mevcc$ to $0.3\mevcc$, and the difference in resolution decreases from $0.3\mevcc$ to $0.1\mevcc$, due to the application of the photon energy correction.
Similarly, for the $\chi_{c2}$ state, the difference in mass is reduced from $2.0\mevcc$ to $0.4\mevcc$ and the resolution improves from $0.7\mevcc$ to $0.4\mevcc$.

\begin{figure}[t]
 \begin{center}
    \includegraphics[width=0.49\textwidth]{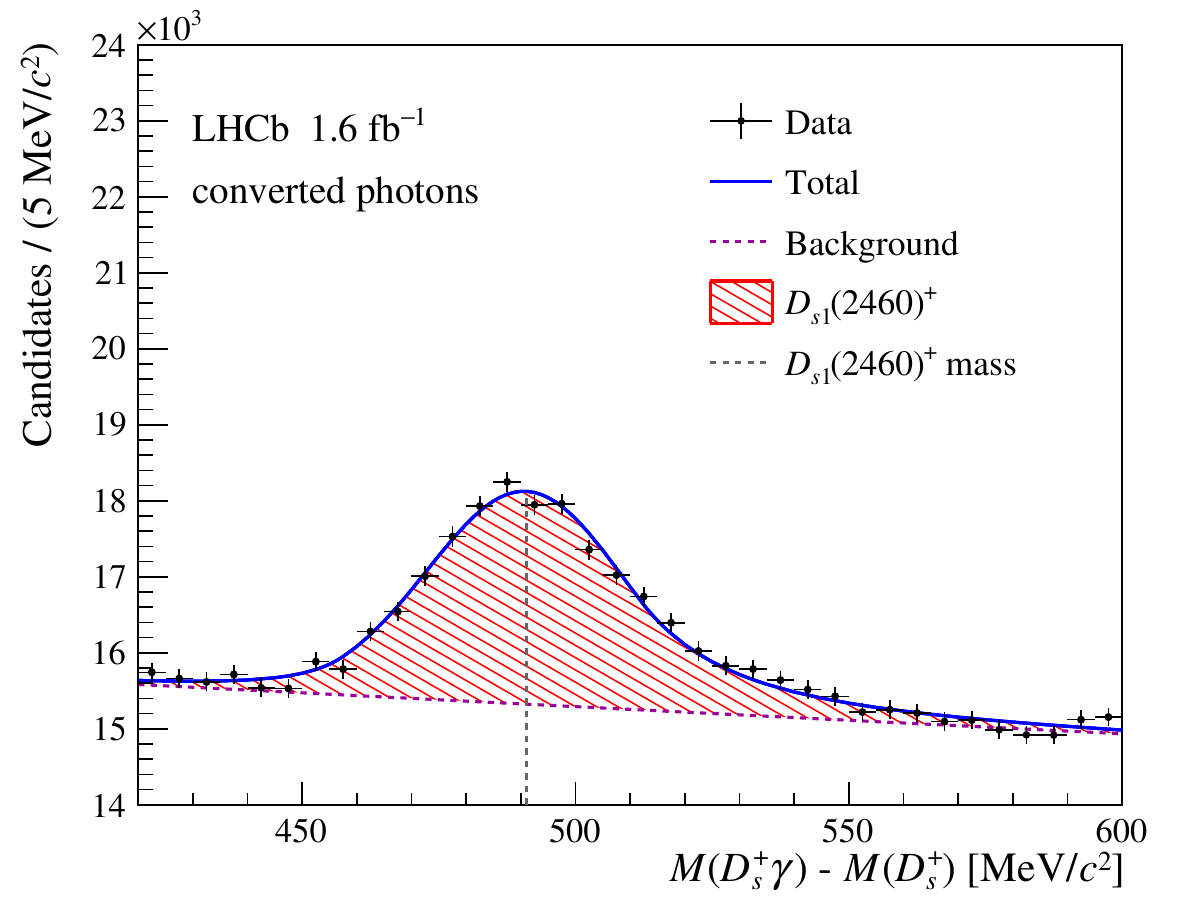}
    \caption{Distribution of $M(D_{s}^{+}\gamma) - M(D_{s}^{+})$ for the selected $D_{s1}(2460)^{+}$ candidates with converted photons. Fit results are also shown. The gray vertical line indicates the known $D_{s1}(2460)^{+}$ mass~\cite{PDG2024}.}
   \label{fig:calib_DsstM}
 \end{center}
\end{figure}

Another channel, $D_{s1}(2460)^+ \to D_s^+ \gamma$,
is used to independently validate the correction method
and assess the associated systematic uncertainty.
The $D_{s1}(2460)^+$ channel has a similar photon energy spectrum and decay topology to the signal channel. The same photon selection criteria as for the $\BcOneP$ candidates are applied.
The $M(\Dsp\gamma) - M(\Dsp)$ distribution for converted photons in 2016 data is shown in Fig.~\ref{fig:calib_DsstM} as an example, along with the result of a fit.
The signal model for $\Dsst$ is a Gaussian function with asymmetric power-law tails,
while the background model is an exponential function.
The peak location and resolution in data are consistently reproduced by the corrected simulation.
Across all the data subsets,
the largest difference remaining in the $D_{s1}(2460)^+$ mass between data and simulation is $2.9\mevcc$, compared to $3.6\mevcc$ before the correction,
while the difference in resolution decreases from $4.6\mevcc$ to $2.6\mevcc$.
The residual difference in the $D_{s1}(2460)^+$ mass is scaled to the signal channel $\BcOnePToBc$ according to the energy released in the decays and taken as the systematic uncertainty arising from the photon energy correction for the $\BcOneP$ mass measurement.

\section{Theory-independent mass fit}
\label{sec:massfit}

As displayed in Fig.~\ref{fig:mass_comp},
a peaking structure is clearly visible in the $M(\Bcp\gamma) - M(\Bcp)$ distribution.
It exhibits a significance above 7 standard deviations when compared to the background-only hypothesis, taking into account the look-elsewhere effect within the predicted mass region~\cite{Gross:2010qma},
as detailed in Ref.~\cite{LHCb-PAPER-2025-014}.
The root mean square of the structure is approximately $62\mevcc$, which largely exceeds the expected value of about $36\mevcc$ for a single peak based on the corrected simulation,
thus indicating the presence of multiple overlapping states.
An unbinned extended maximum-likelihood fit, independent of theoretical inputs, is performed using a minimal two-peak model to describe the observed structure, as detailed in Ref.~\cite{LHCb-PAPER-2025-014}.
The natural widths of the $\BcOneP$ states are expected to be only a few hundred $\kev$~\cite{Eichten:1994gt,Gershtein:1994dxw} and are neglected  due to their small effect compared to the resolution. 
Each peak is modeled by a modified Gaussian function with asymmetric power-law tails as determined from Sec.~\ref{sec:Ecorrection}, leaving only the peak location free to vary in the fit.
The combinatorial background is described by a monotonically increasing third-order polynomial function~\cite{karlin1953geometry}.
The locations of the two peaks are measured to be
\begin{align*}
    M_1 &= 6704.8 \pm 5.5 \pm 2.8 \pm 0.3 \  \mevcc, \\
    M_2 &= 6752.4 \pm 9.5 \pm 3.1 \pm 0.3 \  \mevcc,
\end{align*}
where the uncertainties are statistical, systematic, and from the limited knowledge on the $\Bcp$ mass, given as $6274.5 \pm 0.3 \mevcc$~\cite{PDG2024}.
The statistical correlation between the two measured peak locations is +56\%, while their systematic uncertainties are fully correlated.

\section{$\BcOneP$ production}
\label{sec:production}

Given the current experimental resolution, the two-peak hypothesis serves as the minimal model that effectively describes the data, and extracting more information from the data alone is challenging. Therefore, a theory-constrained model consisting of six peaks is proposed to further explore the properties of the unresolved $\BcOneP$ states.
In this framework, the positions of all six peaks are constrained to the theoretical predictions of $M(\BcOneP)-M(\Bcp)$, whereas those arising from decays with an unreconstructed soft photon from the intermediate $\Bcstar$ state are shifted by $\delta M$ accordingly. 
Their resolutions are determined from the linear dependence derived from simulations generated at different $\BcOneP$ masses.
The production cross section of the $\BcOneP$ states relative to the $\Bcp$ ground state, denoted as $R$, 
is measured in the fiducial region $\pt(\Bcp) < 20 \gevc$ and $2.0 < y(\Bcp) < 4.5$, where $y$ is the rapidity, 
at the center-of-mass energy of $\sqrt{s}=13\tev$, using data corresponding to an integrated luminosity of $6 \, \invfb$.
This subsample is considered to facilitate a direct comparison with theoretical predictions from NRQCD-based calculations~\cite{ 
Eichten:2019gig, 
Chang:2005bf,
Chang:2015qea, Chang:2003cq, Chang:2005hq, Wang:2012ah, Wu:2013pya} at this center-of-mass energy.
The relative production rate is measured as
\begin{align}
\label{eq:prod}
R 
&= \frac{N(\BcOnePToBc)}{N(\Bcp)} \cdot \frac{\varepsilon(\Bcp)}{\varepsilon(\BcOnePToBc)}
\end{align}
where $N(\BcOnePToBc)$ and $N(\Bcp)$ are the signals yields determined from fits to the mass spectra.
The terms $\varepsilon(\Bcp)$ and $\varepsilon(\BcOnePToBc)$ are the corresponding total detection efficiencies for inclusive $\BcToJpsipi$ and $\BcOnePToBc$, respectively, determined using simulation.
With the six $\BcOneP$ peaks predicted from the theory-constrained model,
the detection efficiencies $\varepsilon(\BcOnePToBc)$ are evaluated separately for each corresponding decay process,
with or without the intermediate $\Bcstar$ state.

\begin{figure}[t]
 \begin{center}
    \includegraphics[width=0.5\textwidth]{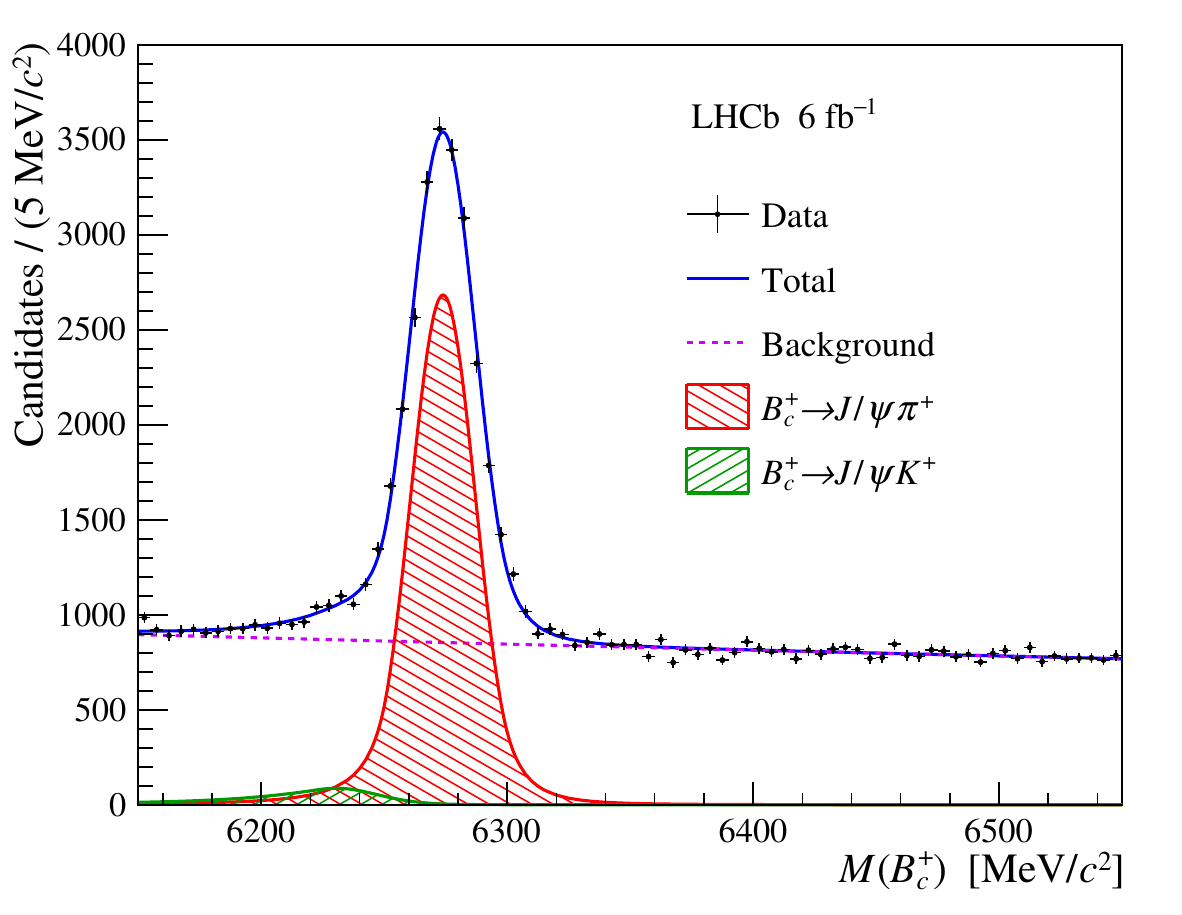}
    \caption{Mass distribution of the selected $B_{c}^{+}$ candidates reconstructed from the $B_{c}^{+} \to J/\psi \pi^+$ decay. The fit results are also shown.}
   \label{fig:massfit_BcM}
 \end{center}
\end{figure}

Figure~\ref{fig:massfit_BcM} shows the mass distribution of the $\Bcp$ candidates,
with the dimuon mass constrained to the known mass of the $\jpsi$ meson~\cite{PDG2024} and the $\Bcp$ candidates required to originate from the associated PV~\cite{Hulsbergen:2005pu}. 
An unbinned extended maximum-likelihood fit is performed on the mass distribution of the selected $\Bcp$ candidates to extract the $\Bcp$ signal yield.
In the fit, both the signal and the background component due to Cabibbo-suppressed decays $\Bcp \to \jpsi K^+$, where the kaon is misidentified as a pion,
are modeled using a modified Gaussian function with asymmetric power-law tails.
The model parameters are fixed from simulation except for the peak location and width of the signal, which are free in the fit, and the peak location of the $\Bcp \to \jpsi K^+$ background, which is constrained relative to that of the signal.
The combinatorial background is described with an exponential function. 
The signal yield of the selected $\Bcp$ candidates is determined to be $(18.70 \pm 0.21)\times 10^3$.

In the theory-constrained fit to the $\Bc\gamma$ candidates, the overall $\BcOneP$ signal shape is defined using theoretical inputs.
A rich set of theoretical predictions on the $\BcOneP$ mass spectrum is taken into account,
as listed in Table~\ref{tab:mass_predictions}.
These predictions are used to fix the location of each peak, while the resolution is determined by the mass-resolution relation described in Sec.~\ref{sec:Ecorrection}.
In addition,
the relative yields of the six signal components are also fixed to stabilize the fit.
The contribution from the $i$ th signal component is given by
\begin{equation}
    N_{i} = \lum \cdot \sigma_{\text{prod},\,i} \cdot \BF_{i} \cdot \varepsilon_i,
\end{equation}
where \lum represents the integrated luminosity and is common to all components, $\sigma_{\text{prod},\,i}$ is the production cross section, $\BF_{i}$ is the decay branching ratio, and $\varepsilon_i$ is the detection efficiency.
The branching ratios $\BF_{i}$ for the $1P_1$ and $1P_1'$ states decaying to $\Bcp \gamma$ or $\Bcstar \gamma$ are calculated based on Ref.~\cite{Godfrey:2004ya},
with the masses and the mixing angle replaced by the values taken from each theoretical model.
The inclusive production cross sections for each of the four $\BcOneP$ states include both direct production, as reported in Table~\ref{tab:prod_predictions}, and feed-down contributions from the decays of excited $B_c(2S)^+$ states. 
The branching fractions for these feed-down processes are calculated according to Ref.~\cite{Godfrey:2004ya}.
The detection efficiency is evaluated using fully simulated samples generated with {\textsc{BcVegPy}}, where the kinematic distributions are corrected to match the various decay modes of the $\BcOneP$ states.
With the peak locations and relative contributions of each component fixed, the overall $\BcOneP$ signal shape is obtained for each theoretical model, allowing the summed $\BcOneP$ signal yield to be determined from the fit.

\begin{figure}[t]
 \begin{center}
    \includegraphics[width=0.49\textwidth]{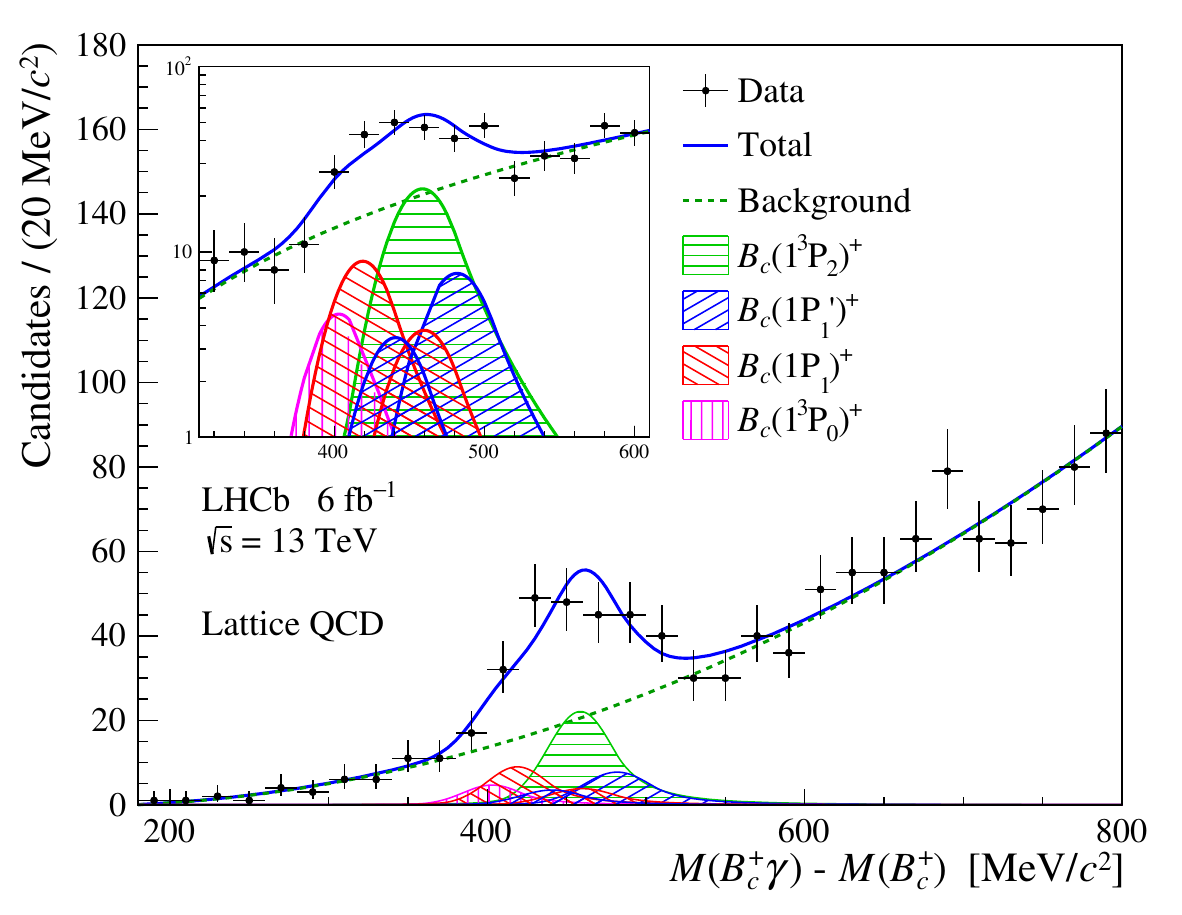}
    \caption{Distribution of the mass difference $M(B_{c}^{+}\gamma)-M(B_{c}^{+})$ with the result of the lattice QCD fit model also shown. 
    The contributions of the six signal peaks from four $B_{c}(1P)^{+}$ states are shown, where both the location and the relative contributions are fixed to the central values predicted by C.T.H. Davies {\it{et al.}}~\cite{Davies:1996gi}. The inset illustrates a zoomed signal region in logarithmic scale. 
    }
   \label{fig:massfit_lattice}
 \end{center}
\end{figure}

\begin{figure}[t]
 \begin{center}
    \includegraphics[width=0.49\textwidth]{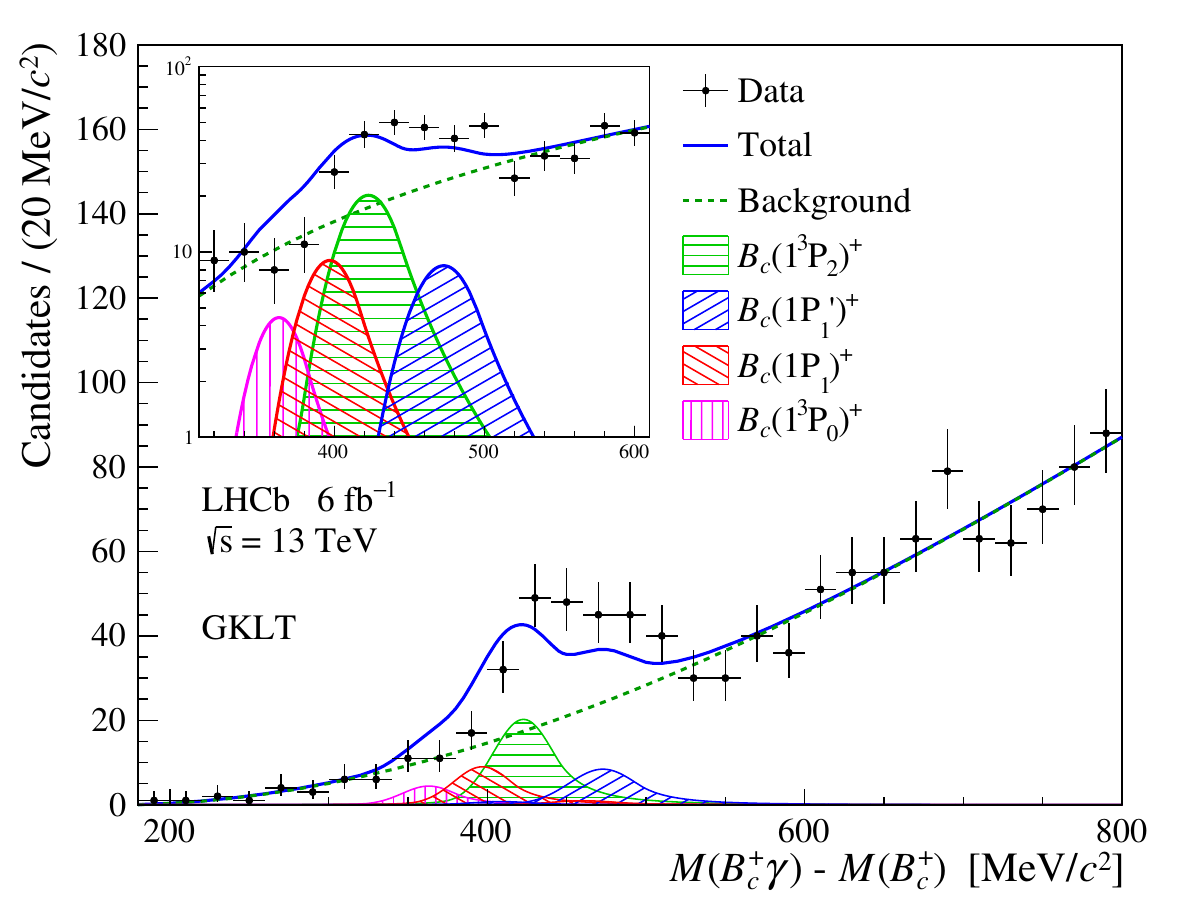}
    \includegraphics[width=0.49\textwidth]{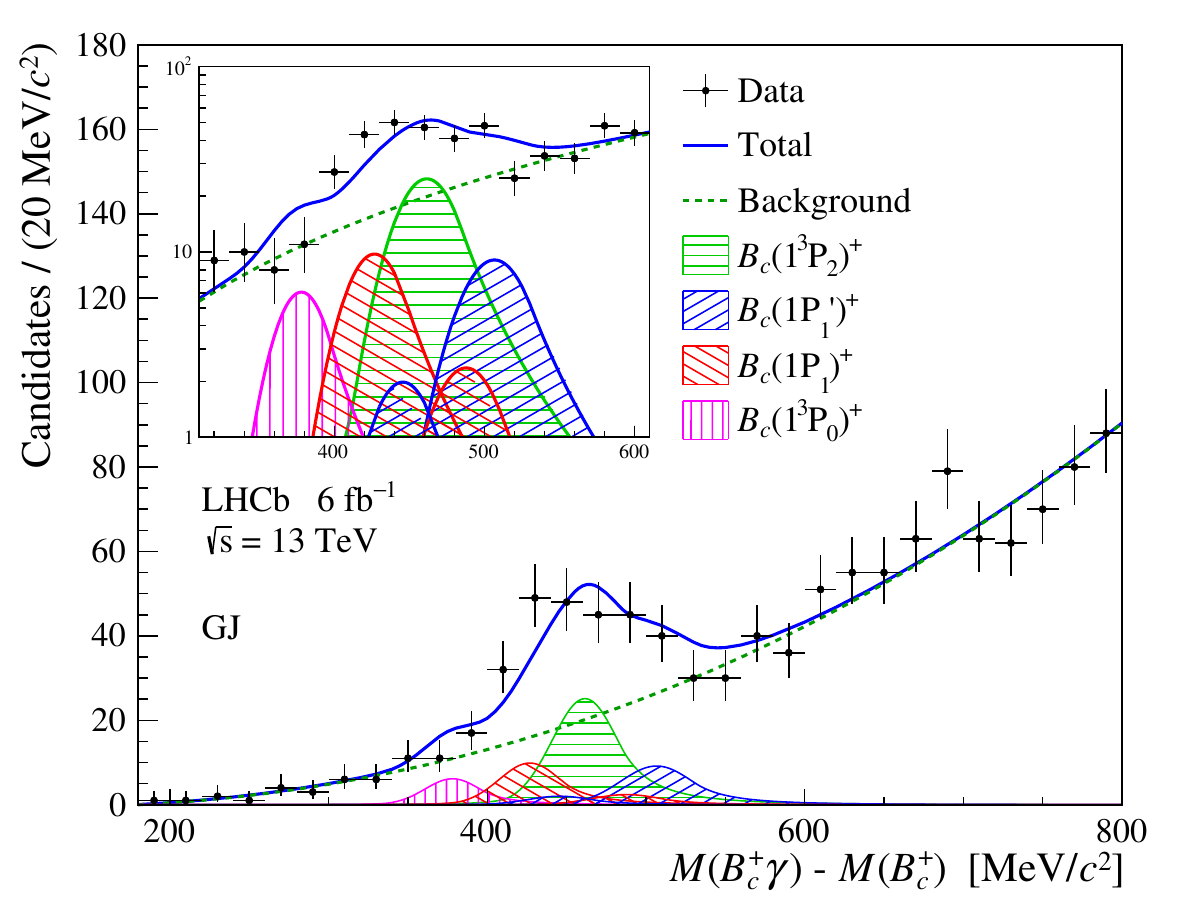}
    \includegraphics[width=0.49\textwidth]{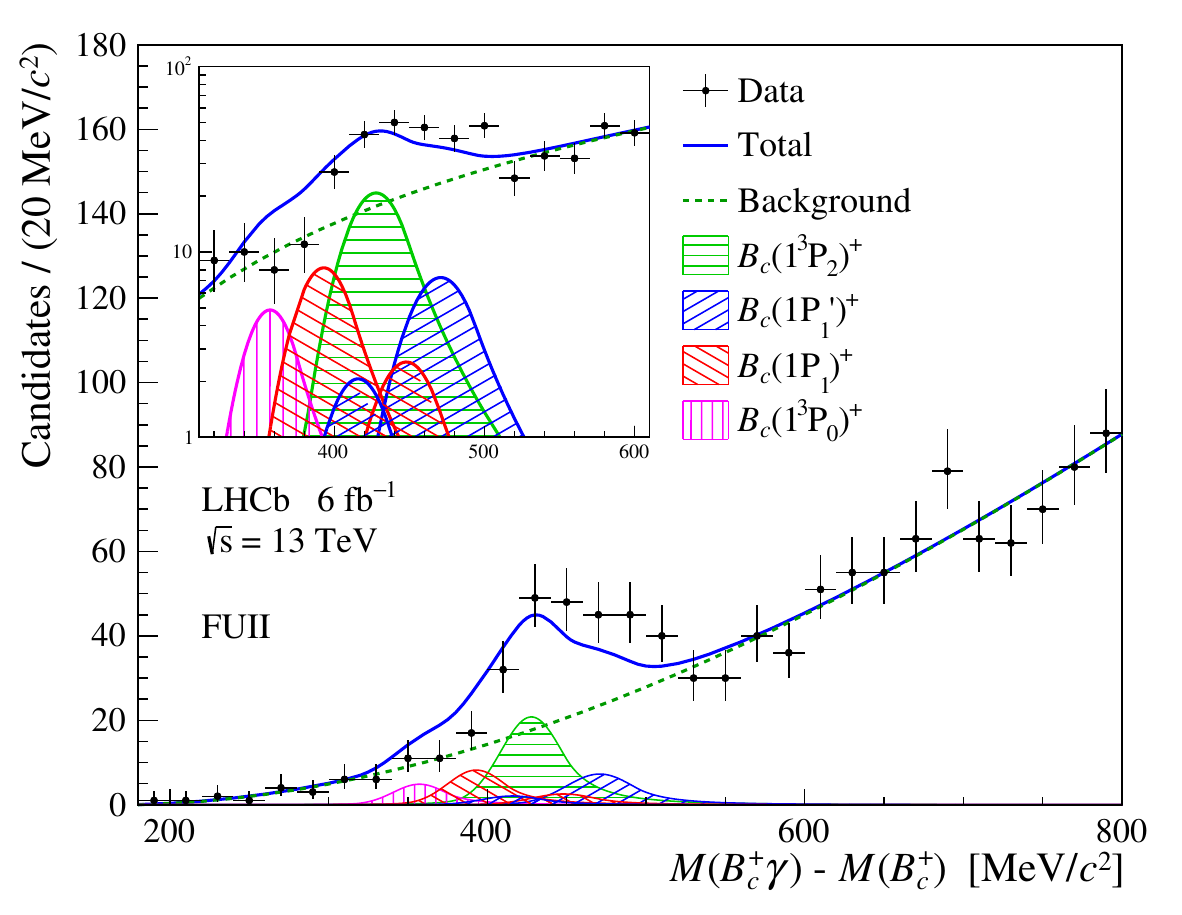}
    \includegraphics[width=0.49\textwidth]{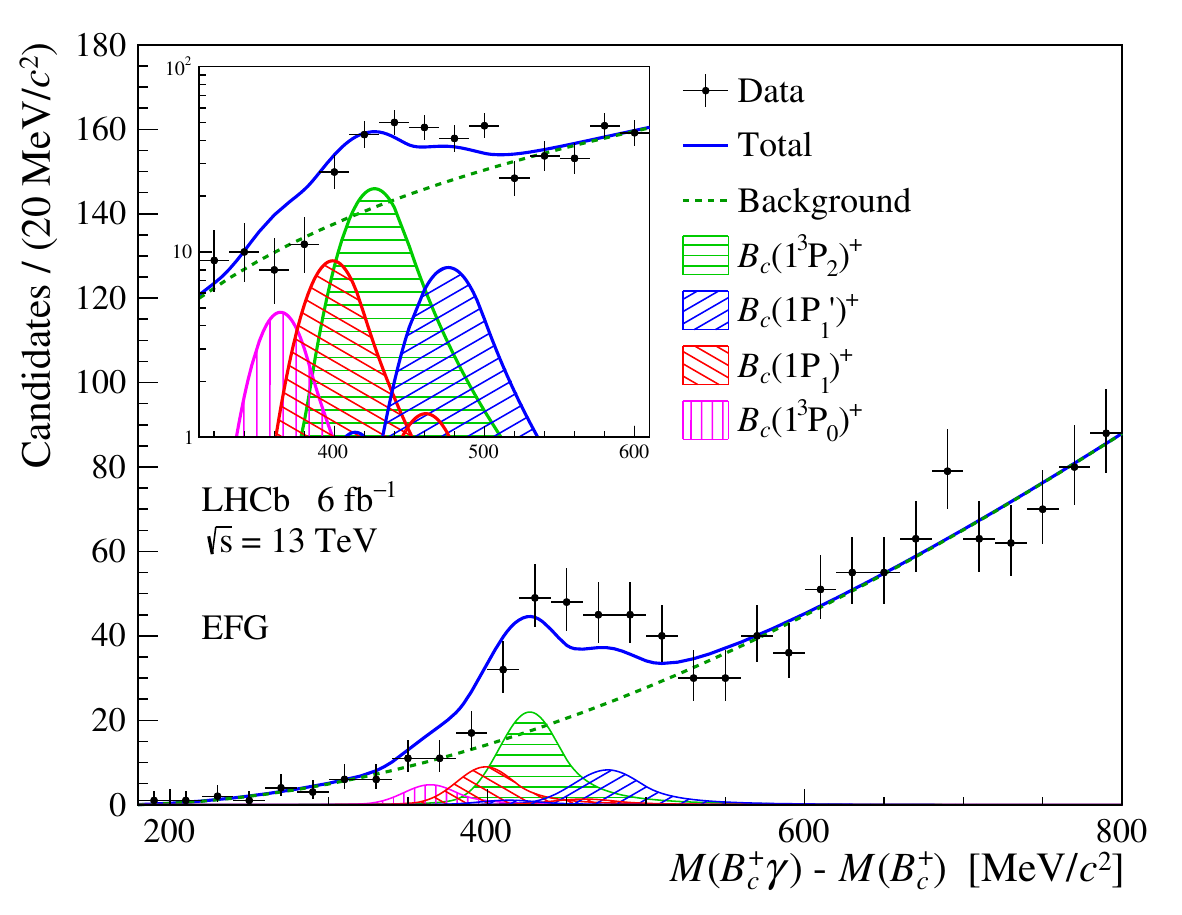}
    \includegraphics[width=0.49\textwidth]{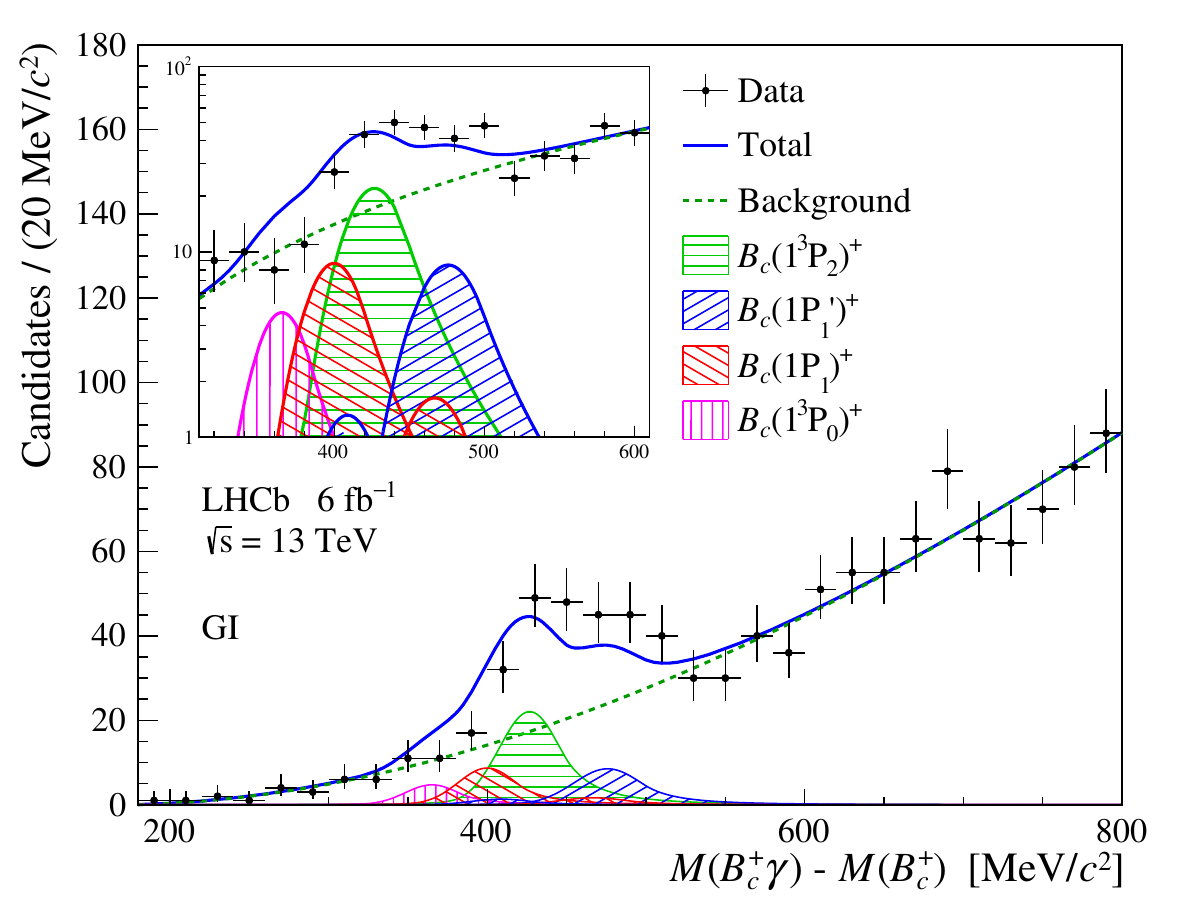}
    \includegraphics[width=0.49\textwidth]{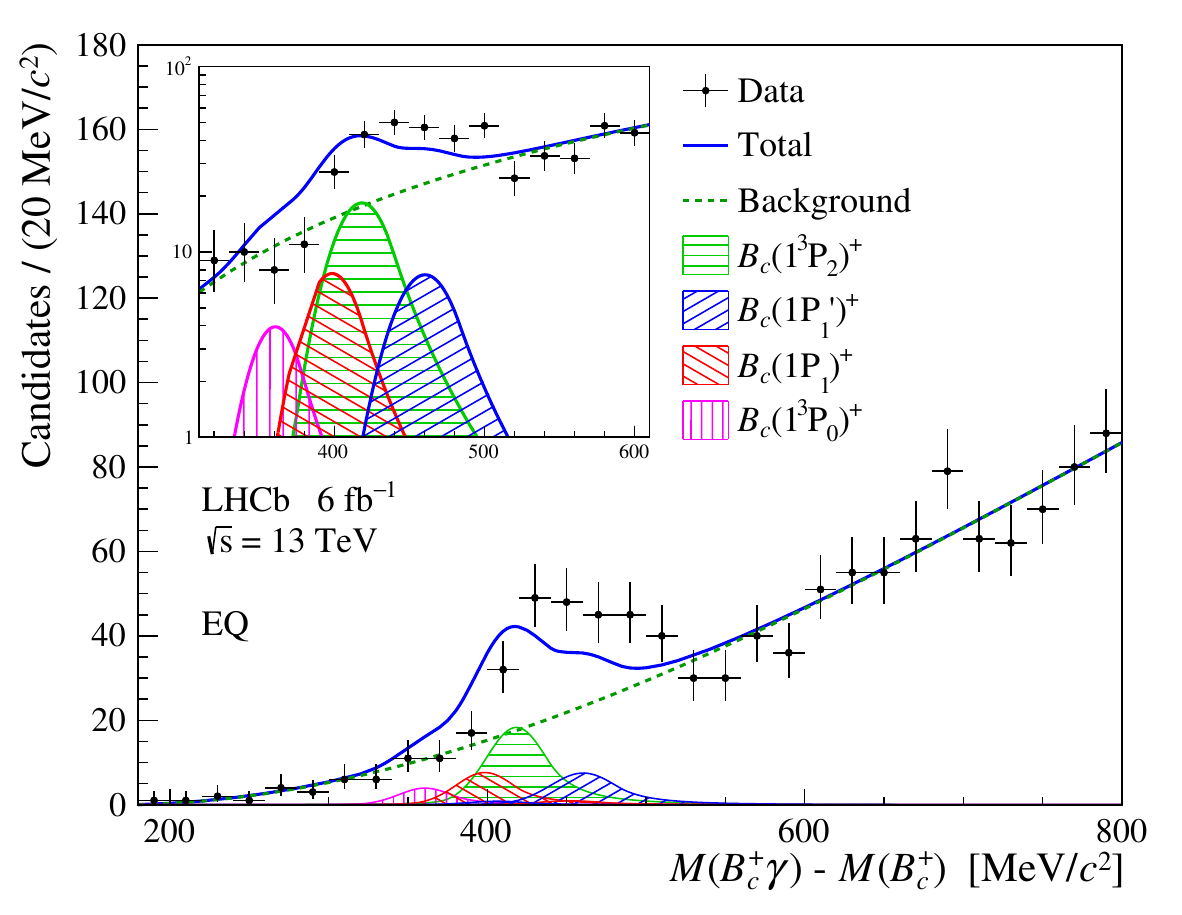}
    \caption{The caption for each plot is the same as Fig.~\ref{fig:massfit_lattice}. Both the location and the relative contributions are fixed to the central value predicted by
    (top left) GKLT~\cite{Gershtein:1994dxw};
    (top right) GJ~\cite{Gupta:1995ps};
    (middle left) FUII~\cite{Fulcher:1998ka};
    (middle right) EFG~\cite{Ebert:2002pp};
    (bottom left) GI~\cite{Godfrey:2004ya};
    (bottom right) EQ~\cite{Eichten:2019gig}.
}
   \label{fig:theo_fit_A}
 \end{center}
\end{figure}

\begin{figure}[t]
 \begin{center}
    \includegraphics[width=0.49\textwidth]{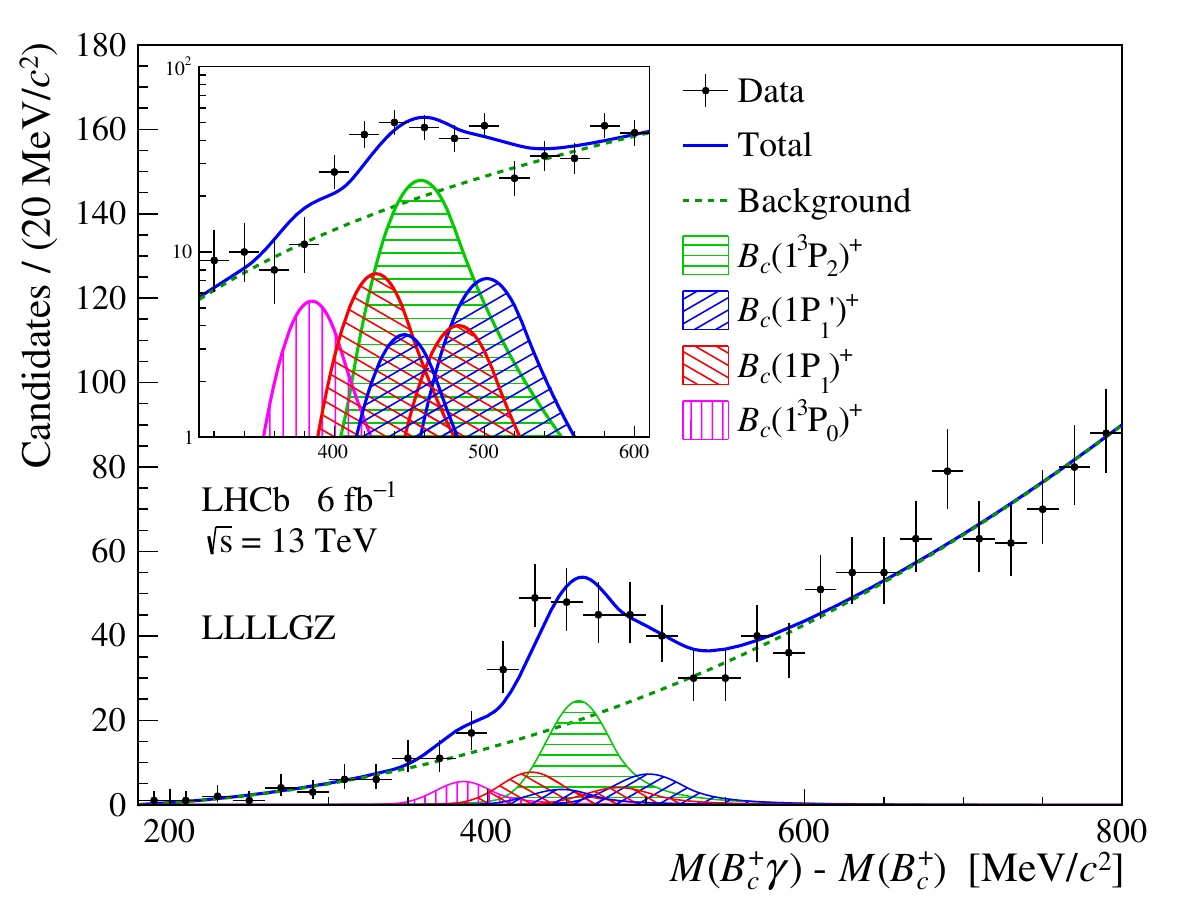}
    \includegraphics[width=0.49\textwidth]{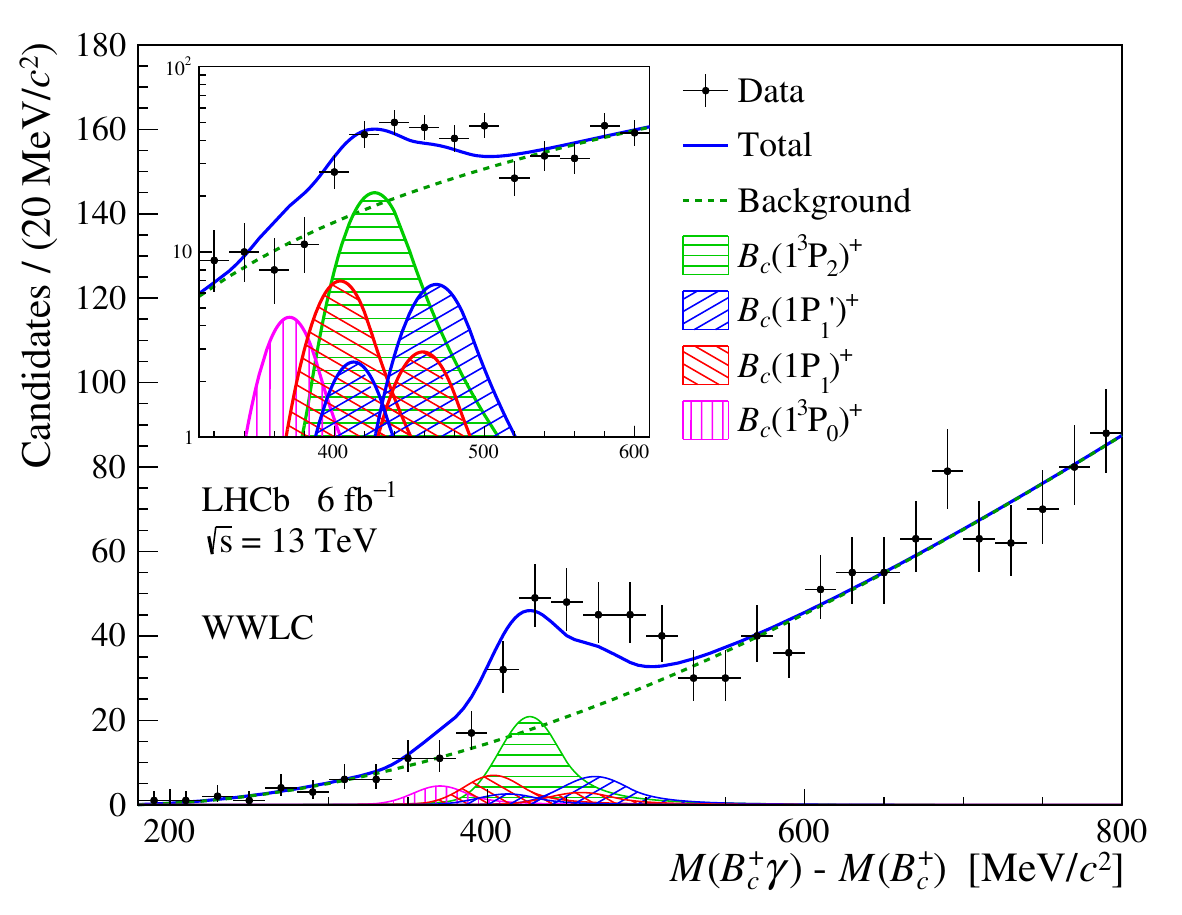}
    \includegraphics[width=0.49\textwidth]{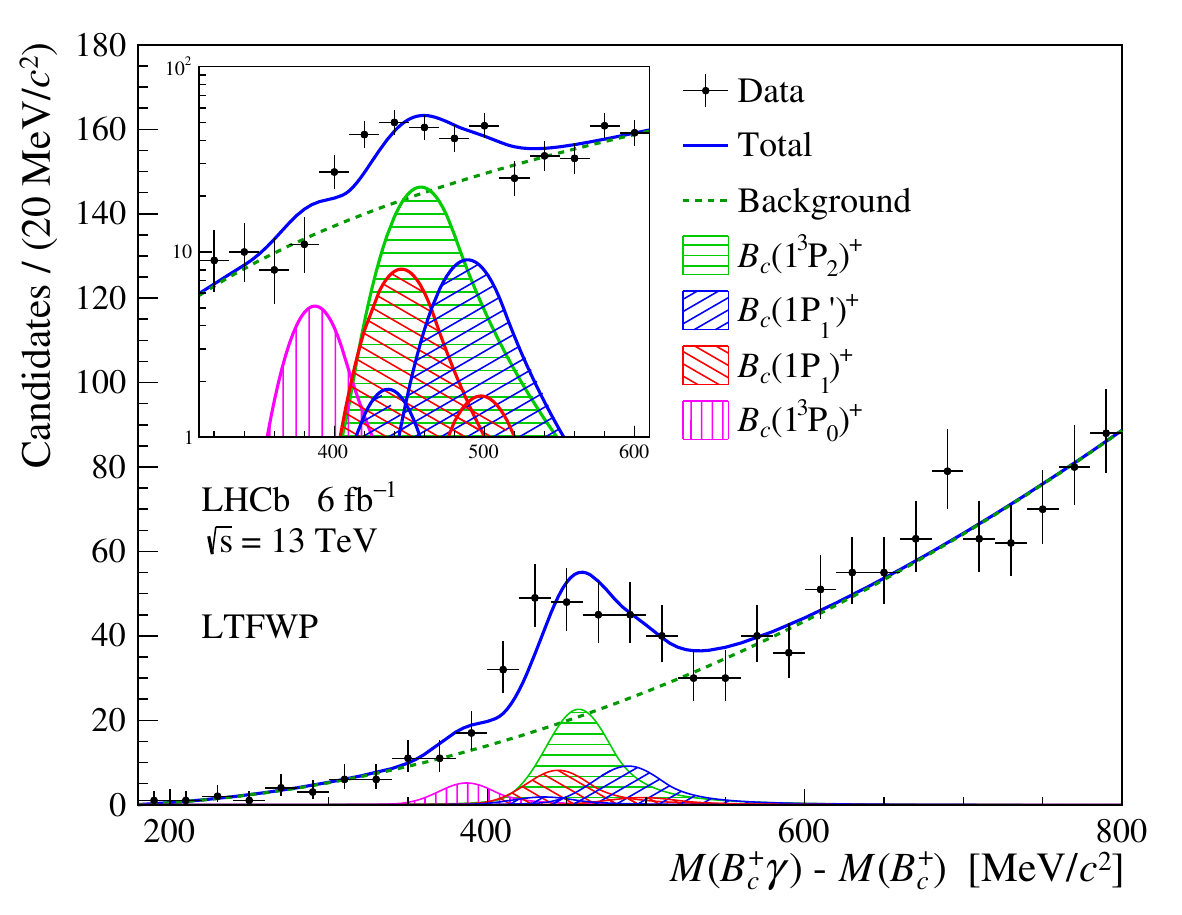}
    \includegraphics[width=0.49\textwidth]{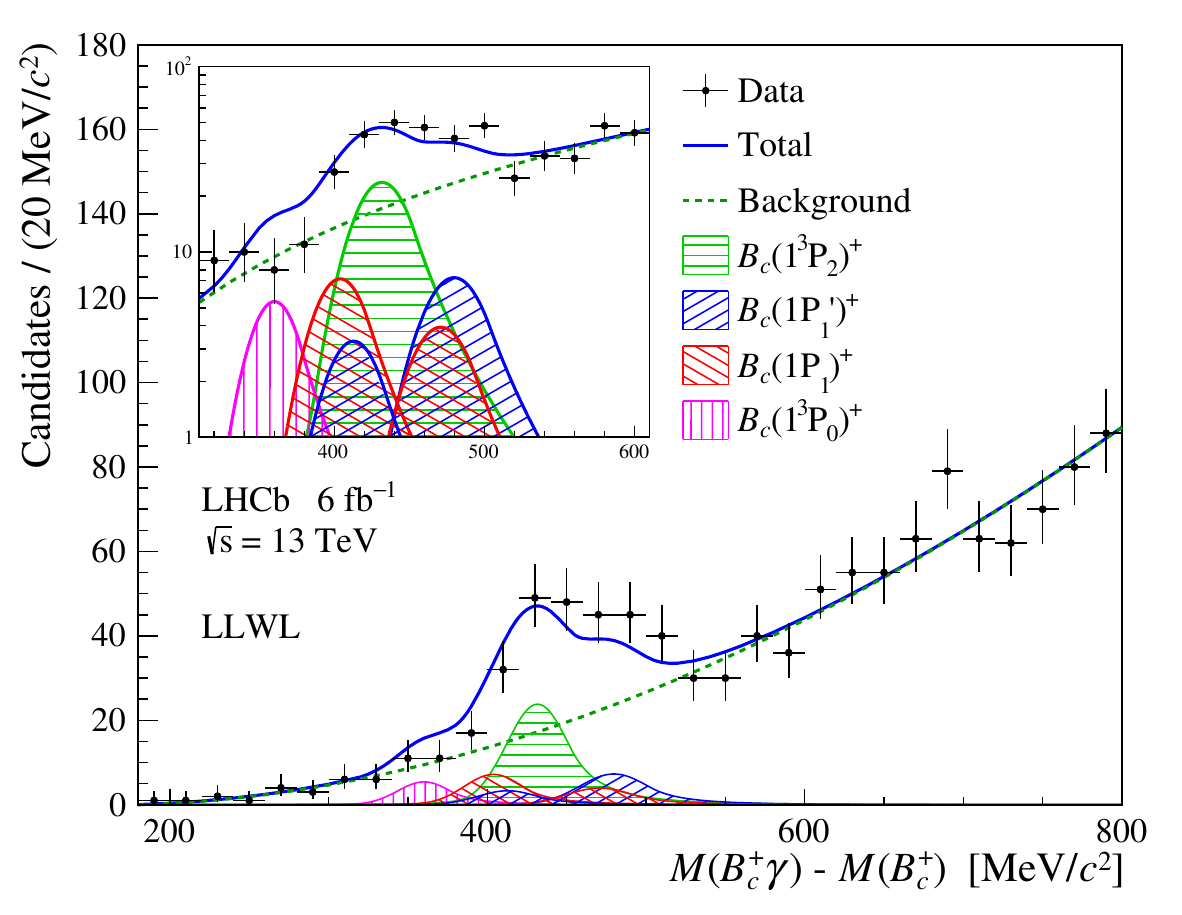}
    \includegraphics[width=0.49\textwidth]{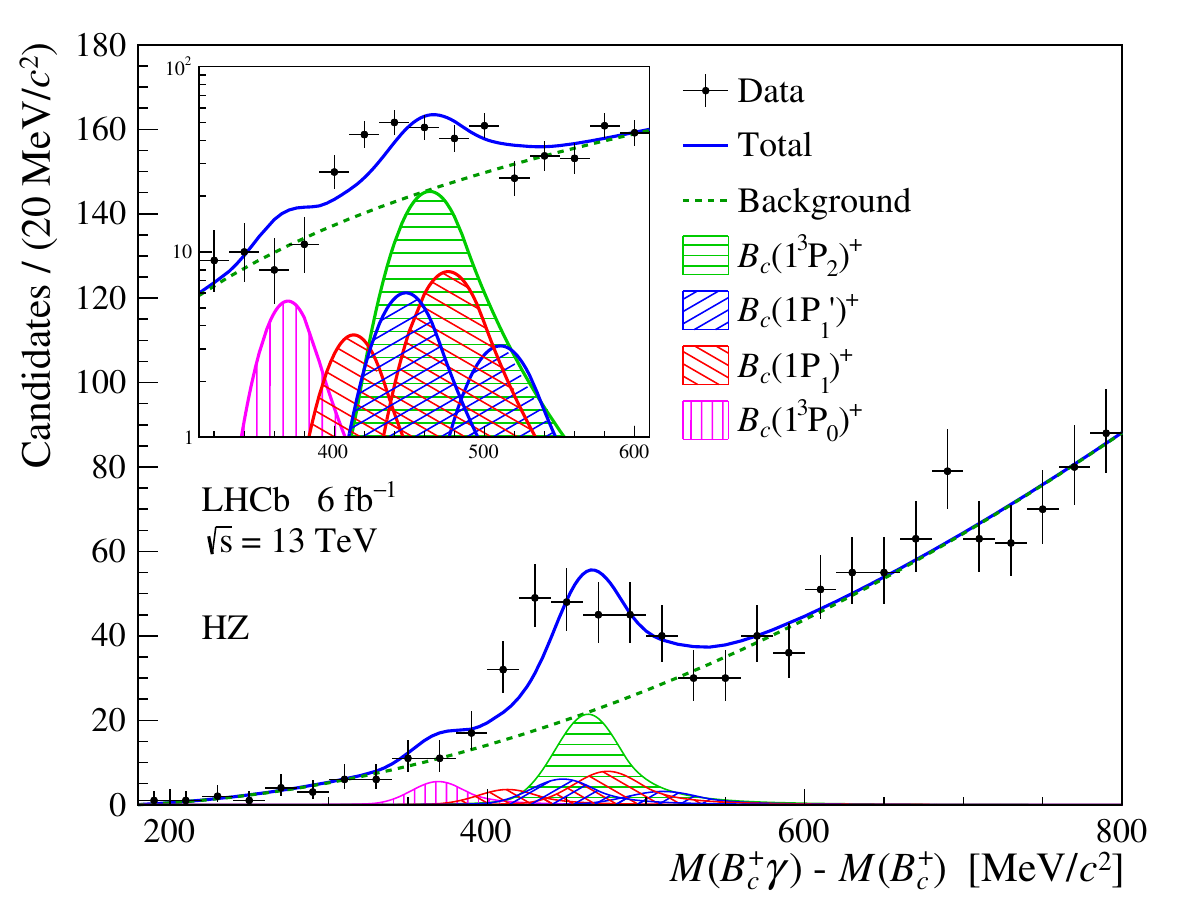}
    \caption{The caption for each plot is the same as Fig.~\ref{fig:massfit_lattice}. Both the location and the relative contributions are fixed to the central value predicted by
    (top left) LLLLGZ~\cite{Li:2019tbn};
    (top right) WWLC~\cite{Wang:2022cxy};
    (middle left) LTFWP~\cite{Li:2022bre};
    (middle right) LLWL~\cite{Li:2023wgq};
    (bottom) HZ~\cite{Hao:2024nqb}.
}
   \label{fig:theo_fit_B}
 \end{center}
\end{figure}

An unbinned extended maximum-likelihood fit to the $M(\Bcp\gamma) - M(\Bcp)$ distribution is performed for each theoretical model,
where the background component is modeled by a third-order polynomial function monotonically increasing in the fit range. 
All the considered models provide a generally good description of the data, with $p$ values ranging from $15\%$ to $90\%$, as estimated using the Kolmogorov--Smirnov criterion~\cite{smirnov1939estimation}. 
The fit results are shown in Figs.~\ref{fig:massfit_lattice} -- \ref{fig:theo_fit_B}.
The fit quality of a given model depends not only on the predicted masses but also on the relative yields of the six signal components. These yields are governed by the predicted production cross sections~\cite{Chang:2015qea} and branching fractions~\cite{Godfrey:2004ya}, the latter of which additionally incorporate the mixing angle as an input parameter.

\begin{table}[!bt]
\centering
\caption{Production cross sections for $\BcOneP$ and $B_c(2S)^+$ states (in unit of nb) calculated using the latest {\textsc{BcVegPy}} 2.2 setup~\cite{Chang:2015qea} for \proton\proton collisions at $\sqs = 13\tev$, in the kinematic region defined by $\pt(\Bcp)<20\ \gevc$ and $2.0 < y(\Bcp) < 4.5$.}
\label{tab:prod_predictions}
\begin{tabular}{c|cccccc}
\toprule
         & $1^1P_1$ & $1^3P_0$ & $1^3P_1$ & $1^3P_2$ & $2^1S_0$ & $2^3S_1$\\ \hline 
 $\sigma_{\text{prod}}$ [nb]      & 3.5 & 1.1 & 2.7 & 7.2 & 4.5 & 11.1   \\
\bottomrule
\end{tabular}
\end{table}

The fit shown in Fig.~\ref{fig:massfit_lattice} is selected as the baseline for the production measurement as it provides the best $p$ value. It is based on the lattice QCD calculations of Ref.~\cite{Davies:1996gi}. 
The signal yield of $\BcOneP$ is determined to be $153 \pm 22$, including only the statistical uncertainty.
It is further corrected for a tiny~($1.3\%$) contribution from misidentified $\Bcp \to \jpsi K^+$ candidates.
The theoretical uncertainty is evaluated as the largest difference between the baseline result and those obtained from alternative QCD models that are consistent with the data within one standard deviation~\cite{Godfrey:2004ya, 
Ebert:2002pp,
Fulcher:1998ka,
Gupta:1995ps,
Wang:2022cxy,
Li:2023wgq,
Hao:2024nqb,
Li:2019tbn,
Li:2022bre}.
The largest deviation is assigned as the theoretical model uncertainty.
The relative production rate is measured to be
\begin{equation*}
    R = 0.20 \pm 0.03  \pm 0.03,
\end{equation*}
where the terms denote statistical and theoretical uncertainties, respectively.
The result is consistent with NRQCD-based theoretical predictions~\cite{Godfrey:2004ya,Eichten:2019gig}. These predictions, calculated with {\textsc{BcVegPy}}, yield values from 0.17 to 0.19, including color-octet contributions and feed-down from $B_c(2S)^+$ decays.

\FloatBarrier

\section{Systematic uncertainties on production measurement}
\label{sec:systematics}

Several sources of systematic uncertainty that affect the measurement of the relative production rate are considered. 
Their contributions are summarized in Table~\ref{tab:unc_prod} and treated as uncorrelated. 
The imperfect modeling of the signal and background shapes can affect the $\BcOneP$ and $\Bcp$ yields, biasing the final result. 
Therefore, alternative fit models are used to assess their potential impact.
The signal shapes of $\Bcp$ and each $\BcOneP$ components are replaced with a sum of the original modified Gaussian function with asymmetric power-law tails and an additional Gaussian function,
leading to a systematic uncertainty of $0.8\%$.
For the $\Bcp$ background shape, the exponential function is replaced with a second-order polynomial function, while the background for $\BcOneP$ is described using a monotonically increasing fourth-order polynomial function instead.
The difference with respect to the nominal result is $6.9\%$,
taken as the corresponding systematic uncertainty.

In the efficiency calculation, the photon detection efficiency is corrected using results from Ref.~\cite{Govorkova:2015vqa}. 
The correction factors are determined from the relative yields of $B^+ \to \jpsi K^{*+} (\to K^+ \pi^0)$ versus $B^+ \to \jpsi K^+$ events.
The uncertainty in the correction factors, primarily due to the branching ratio of \mbox{$B^+ \to \jpsi K^{*+}$}~\cite{PDG2024}, is propagated through the efficiency evaluation and included as a systematic uncertainty of $3.5\%$ in the final results.
The systematic uncertainty due to the limited size of the simulated samples is well below 1\%.
The imperfect simulation of particle identification, track reconstruction, and trigger largely cancels in the relative production rate. Their impact is estimated to be of 1\% each.
The systematic uncertainties are in addition to the contribution of 14\% due to the choice of the theoretical model, which is accounted for separately.

\begin{table}[t]
    \centering
    \caption{Summary of relative systematic uncertainties on the measurement of the production rate between $\BcOneP$ and $\Bcp$ states. The total uncertainty is calculated as the sum in quadrature of individual contributions.}
    \label{tab:unc_prod}
    \begin{tabular}{l|c}
    \toprule
    Source &  $\delta R / R$ [\%] \\
    \midrule
    Background shape                 & 6.9   \\
    Signal shape                     & 0.8   \\
    Photon detection                 & 3.5   \\
    Simulation sample size           & 0.6   \\
    Track reconstruction             & 1.0   \\
    PID requirement                  & 1.0   \\
    Trigger requirement              & 1.0   \\
    \midrule
    Total         & 8.0   \\
    \bottomrule
    \end{tabular}
\end{table}

\section{Conclusion}
\label{sec:summary}

In summary, using $pp$ collision data corresponding to an integrated luminosity of $9\,\mathrm{fb}^{-1}$ collected by the LHCb experiment at center-of-mass energies of 7, 8 and 13\,TeV, 
a wide peaking structure is observed in the $\BcGamma$ mass spectrum with a significance larger than 7 standard deviations.
The structure is compatible with the lowest $P$-wave excited states of the beauty-charm meson, as
verified by fits based on the theoretical predictions from lattice QCD~\cite{Davies:1996gi} and various QCD models~\cite{
Godfrey:2004ya, 
Ebert:2002pp,
Fulcher:1998ka,
Gershtein:1994dxw,
Gupta:1995ps,
Wang:2022cxy,
Li:2023wgq,
Hao:2024nqb,
Li:2019tbn,
Li:2022bre,
Eichten:2019gig}.
The structure is well described by a minimal two-peak model in a theory-independent approach,
with their measured masses reported in Ref.~\cite{LHCb-PAPER-2025-014}.
The distinction between contributions from each $\BcOneP$ state requires larger statistics and significant improvements in mass resolution, potentially achieved by using photons that convert in the detector material and are reconstructed as $e^+e^-$ pairs~\cite{LHCb-PAPER-2013-028,LHCb-PAPER-2014-040},
which may enable a measurement of the $\Bcstar$ and $\Bcp$ mass difference.
Such a measurement is of significant theoretical interest and would be challenging to perform otherwise.

Using a subset of data collected at $\sqrt{s}=13 \tev$, the fraction of $\Bcp$ mesons from $\BcOneP$ decays is measured
in the $\Bcp$ kinematic region $\pt < 20 \gevc$ and $2.0 < y < 4.5$. 
It is determined to be 
\begin{equation*}
    R = 0.20 \pm 0.03 \pm 0.02 \pm 0.03,
\end{equation*}
where the first, second, and third uncertainties are statistical, systematic, and theoretical, respectively.
The measured value is in good agreement with expectations from NRQCD-based calculations.
~These results represent the first observation of orbitally excited beauty-charm states and measurement of their production properties, providing crucial insights into the internal dynamics of hadrons containing two heavy quarks and the nonperturbative aspects of QCD.

\emph{Data availability:} Data associated to the plots in this publication are made available on the CERN Document Server in Ref.~\cite{LHCb-PAPER-2025-015-cds}.

\section*{Acknowledgements}
%
%
\noindent
We thank Xing-Gang Wu for valuable discussions on the $\Bcp$ production issues.
We express our gratitude to our colleagues in the CERN
accelerator departments for the excellent performance of the LHC. We
thank the technical and administrative staff at the LHCb
institutes.
We acknowledge support from CERN and from the national agencies:
ARC (Australia);
CAPES, CNPq, FAPERJ and FINEP (Brazil); 
MOST and NSFC (China); 
CNRS/IN2P3 (France); 
BMBF, DFG and MPG (Germany); 
INFN (Italy); 
NWO (Netherlands); 
MNiSW and NCN (Poland); 
MCID/IFA (Romania); 
MICIU and AEI (Spain);
SNSF and SER (Switzerland); 
NASU (Ukraine); 
STFC (United Kingdom); 
DOE NP and NSF (USA).
We acknowledge the computing resources that are provided by ARDC (Australia), 
CBPF (Brazil),
CERN, 
IHEP and LZU (China),
IN2P3 (France), 
KIT and DESY (Germany), 
INFN (Italy), 
SURF (Netherlands),
Polish WLCG (Poland),
IFIN-HH (Romania), 
PIC (Spain), CSCS (Switzerland), 
and GridPP (United Kingdom).
We are indebted to the communities behind the multiple open-source
software packages on which we depend.
Individual groups or members have received support from
Key Research Program of Frontier Sciences of CAS, CAS PIFI, CAS CCEPP, 
Fundamental Research Funds for the Central Universities,  and Sci.\ \& Tech.\ Program of Guangzhou (China);
Minciencias (Colombia);
EPLANET, Marie Sk\l{}odowska-Curie Actions, ERC and NextGenerationEU (European Union);
A*MIDEX, ANR, IPhU and Labex P2IO, and R\'{e}gion Auvergne-Rh\^{o}ne-Alpes (France);
Alexander-von-Humboldt Foundation (Germany);
ICSC (Italy); 
Severo Ochoa and Mar\'ia de Maeztu Units of Excellence, GVA, XuntaGal, GENCAT, InTalent-Inditex and Prog.~Atracci\'on Talento CM (Spain);
SRC (Sweden);
the Leverhulme Trust, the Royal Society and UKRI (United Kingdom).

\clearpage 
\addcontentsline{toc}{section}{References}
\bibliographystyle{LHCb}
\bibliography{main,standard,LHCb-PAPER,LHCb-CONF,LHCb-DP,LHCb-TDR}

\clearpage 
\centerline
{\large\bf LHCb collaboration}
\begin
{flushleft}
\small
R.~Aaij$^{38}$\lhcborcid{0000-0003-0533-1952},
A.S.W.~Abdelmotteleb$^{57}$\lhcborcid{0000-0001-7905-0542},
C.~Abellan~Beteta$^{51}$\lhcborcid{0009-0009-0869-6798},
F.~Abudin{\'e}n$^{57}$\lhcborcid{0000-0002-6737-3528},
T.~Ackernley$^{61}$\lhcborcid{0000-0002-5951-3498},
A. A. ~Adefisoye$^{69}$\lhcborcid{0000-0003-2448-1550},
B.~Adeva$^{47}$\lhcborcid{0000-0001-9756-3712},
M.~Adinolfi$^{55}$\lhcborcid{0000-0002-1326-1264},
P.~Adlarson$^{84}$\lhcborcid{0000-0001-6280-3851},
C.~Agapopoulou$^{14}$\lhcborcid{0000-0002-2368-0147}
% ...

{\footnotesize \it

% ...
$^{14}$Universit{\'e} Paris-Saclay, CNRS/IN2P3, IJCLab, Orsay, France\\
% ...
$^{38}$Nikhef National Institute for Subatomic Physics, Amsterdam, Netherlands\\
% ...
$^{47}$Instituto Galego de F{\'\i}sica de Altas Enerx{\'\i}as (IGFAE), Universidade de Santiago de Compostela, Santiago de Compostela, Spain\\
% ...
$^{51}$Physik-Institut, Universit{\"a}t Z{\"u}rich, Z{\"u}rich, Switzerland\\
% ...
$^{55}$H.H. Wills Physics Laboratory, University of Bristol, Bristol, United Kingdom\\
% ...
$^{57}$Department of Physics, University of Warwick, Coventry, United Kingdom\\
% ...
$^{60}$School of Physics and Astronomy, University of Glasgow, Glasgow, United Kingdom\\
$^{61}$Oliver Lodge Laboratory, University of Liverpool, Liverpool, United Kingdom\\
% ...
$^{69}$Syracuse University, Syracuse, NY, United States\\
% ...
$^{84}$Department of Physics and Astronomy, Uppsala University, Uppsala, Sweden, associated to $^{60}$\\
% ...
}
\end{flushleft}


\end{document}